\documentclass[twocolumn,trackchanges,tighten]{aastex63}

\newcommand{\ha}{H$\alpha$}
\newcommand{\hb}{H$\beta$}
\newcommand{\hd}{H$\delta$}
\newcommand{\kms}{\,km\,s$^{-1}$}
\newcommand{\rsun}{\,R$_{\sun}$}
\newcommand{\lsun}{\,L$_{\sun}$}
\newcommand{\msun}{\,M$_{\sun}$}
\newcommand{\ang}{\,\AA}
\newcommand{\ergs}{\,erg\,s$^{-1}$}
\newcommand{\msunyr}{\,M$_{\sun}$\,yr$^{-1}$}
\newcommand{\mdot}{$\dot{M}$}
\renewcommand{\object}{SN~2017ahn}

\received{2020 August 14}
\revised{2020 November 11}
\accepted{2020 November 12}
\submitjournal{ApJ}

\shorttitle{The Type II SN~2017ahn}
\shortauthors{Tartaglia et al.}

\begin{document}

\title{The early discovery of SN~2017ahn: signatures of persistent interaction in a fast declining Type II supernova}

\correspondingauthor{Leonardo Tartaglia}
\email{leonardo.tartaglia@inaf.it}

\author[0000-0003-3433-1492]{L. Tartaglia}
\affiliation{INAF - Osservatorio Astronomico di Padova, Vicolo dell'Osservatorio 5, 35122 Padova, Italy}
\affiliation{Department of Astronomy and the Oskar Klein Centre, Stockholm University, AlbaNova, Roslagstullsbacken 21, 114 21 Stockholm, Sweden}

\author[0000-0003-4102-380X]{D. J. Sand}
\affiliation{Steward Observatory, University of Arizona, 933 North Cherry Avenue, Rm. N204, Tucson, AZ 85721-0065, USA}

\author[0000-0001-7675-3381]{J. H. Groh}
\affiliation{School of Physics, Trinity College Dublin, the University of Dublin, Dublin, Ireland}

\author[0000-0001-8818-0795]{S. Valenti}
\affiliation{Department of Physics, University of California, 1 Shields Avenue, Davis, CA 95616-5270, USA}

\author[0000-0003-2732-4956]{S. D. Wyatt}
\affiliation{Steward Observatory, University of Arizona, 933 North Cherry Avenue, Rm. N204, Tucson, AZ 85721-0065, USA}

\author[0000-0002-4924-444X]{K. A. Bostroem}
\affiliation{Department of Physics, University of California, 1 Shields Avenue, Davis, CA 95616-5270, USA}

\author[0000-0001-6272-5507]{P. J. Brown}
\affiliation{Mitchell Institute for Fundamental Physics and Astronomy, Texas A\&M University,College Station, TX 77843, USA}

\author[0000-0002-2898-6532]{S. Yang}
\affiliation{Department of Astronomy and the Oskar Klein Centre, Stockholm University, AlbaNova, Roslagstullsbacken 21, 114 21 Stockholm, Sweden}

\author{J. Burke}
\affiliation{Las Cumbres Observatory, 6740 Cortona Drive, Suite 102, Goleta, CA 93117-5575, USA}
\affiliation{Department of Physics, University of California, Santa Barbara, CA 93106-9530}

\author[0000-0003-1532-0149]{T. -W. Chen}
\affiliation{Department of Astronomy and the Oskar Klein Centre, Stockholm University, AlbaNova, Roslagstullsbacken 21, 114 21 Stockholm, Sweden}
\affiliation{Max-Planck-Institut f{\"u}r Extraterrestrische Physik, Giessenbachstra\ss e 1, 85748, Garching, Germany}

\author[0000-0002-2806-5821]{S. Davis}
\affiliation{Department of Physics, Florida State University, 77 Chieftan Way, Tallahassee, FL 32306, USA}

\author[0000-0003-3459-2270]{F. F\"orster}
\affiliation{Millennium Institute of Astrophysics (MAS), Nuncio Monse\~nor Sotero Sanz 100, Providencia, Santiago, Chile}
\affiliation{Center for Mathematical Modelling, Universidad de Chile, Avenida Blanco Encalada 2120 Piso 7, Santiago, Chile}

\author[0000-0002-1296-6887]{L. Galbany}
\affiliation{Departamento de F\'isica Te\'orica y del Cosmos, Universidad de Granada, E-18071 Granada, Spain}

\author{J. Haislip}
\affiliation{Department of Physics and Astronomy, University of North Carolina at Chapel Hill, Chapel Hill, NC 27599, USA}

\author[0000-0002-1125-9187]{D. Hiramatsu}
\affiliation{Las Cumbres Observatory, 6740 Cortona Drive, Suite 102, Goleta, CA 93117-5575, USA}
\affiliation{Department of Physics, University of California, Santa Barbara, CA 93106-9530}

\author[0000-0002-0832-2974]{G. Hosseinzadeh}
\affiliation{Center for Astrophysics \textbar{} Harvard \& Smithsonian, 60 Garden Street, Cambridge, MA 02138-1516, USA}

\author[0000-0003-4253-656X]{D. A. Howell}
\affiliation{Las Cumbres Observatory, 6740 Cortona Drive, Suite 102, Goleta, CA 93117-5575, USA}
\affiliation{Department of Physics, University of California, Santa Barbara, CA 93106-9530}

\author[0000-0003-1039-2928]{E. Y. Hsiao}
\affiliation{Department of Physics, Florida State University, 77 Chieftan Way, Tallahassee, FL 32306, USA}

\author[0000-0001-8738-6011]{S. W. Jha}
\affiliation{Department of Physics and Astronomy, Rutgers the State University of New Jersey, 136 Frelinghuysen Road, Piscataway, NJ 08854, USA}

\author[0000-0003-3642-5484]{V. Kouprianov}
\affiliation{Department of Physics and Astronomy, University of North Carolina at Chapel Hill, Chapel Hill, NC 27599, USA}

\author[0000-0002-1132-1366]{H. Kuncarayakti}
\affiliation{Tuorla Observatory, Department of Physics and Astronomy, FI-20014 University of Turku, Finland}
\affiliation{Finnish Centre for Astronomy with ESO (FINCA), FI-20014 University of Turku, Finland}

\author{J. D. Lyman}
\affiliation{Department of Physics, University of Warwick, Coventry CV4 7AL, UK}

\author[0000-0001-5807-7893]{C. McCully}
\affiliation{Las Cumbres Observatory, 6740 Cortona Drive, Suite 102, Goleta, CA 93117-5575, USA}

\author[0000-0003-2734-0796]{M. M. Phillips}
\affiliation{Carnegie Observatories, Las Campanas Observatory, Casilla 601, La Serena, Chile}

\author[0000-0001-5990-6243]{A. Rau}
\affiliation{Max-Planck-Institut f{\"u}r Extraterrestrische Physik, Giessenbachstra\ss e 1, 85748, Garching, Germany}

\author[0000-0002-5060-3673]{D. E. Reichart}
\affiliation{Department of Physics and Astronomy, University of North Carolina at Chapel Hill, Chapel Hill, NC 27599, USA}

\author[0000-0002-9301-5302]{M. Shahbandeh}
\affiliation{Department of Physics, Florida State University, 77 Chieftan Way, Tallahassee, FL 32306, USA}

\author[0000-0002-1468-9668]{J. Strader}
\affiliation{Center for Data Intensive and Time Domain Astronomy, Department of Physics and Astronomy, Michigan State University, East Lansing, MI 48824, USA}

\begin{abstract}
We present high-cadence, comprehensive data on the nearby ($D\simeq33\,\rm{Mpc}$) Type II \object, discovered within $\sim$1 day of explosion, from the very early phases after explosion to the nebular phase.
The observables of \object~show a significant evolution over the $\simeq470\,\rm{d}$ of our follow-up campaign, first showing prominent, narrow Balmer lines and other high-ionization features purely in emission (i.e. flash spectroscopy features), which progressively fade and lead to a spectroscopic evolution similar to that of more canonical Type II supernovae.
Over the same period, the decline of the light curves in all bands is fast, resembling the photometric evolution of linearly declining H-rich core-collapse supernovae.
The modeling of the light curves and early flash spectra suggest a complex circumstellar medium surrounding the progenitor star at the time of explosion, with a first dense shell produced during the very late stages of its evolution being swept up by the rapidly expanding ejecta within the first $\sim6\,\rm{d}$ of the supernova evolution, while signatures of interaction are observed also at later phases.
Hydrodynamical models support the scenario in which linearly declining Type II supernovae are predicted to arise from massive yellow super/hyper giants depleted of most of their hydrogen layers.
\end{abstract}


\section{Introduction} \label{sec:intro}
Core-collapse supernovae (CC SNe) are the spectacular endpoint of the evolution of massive stars \citep[$>8-9$\msun;][]{2003ApJ...591..288H,2009ARA&A..47...63S}.
Hydrogen-rich SNe are typically labelled as Type II SNe \citep{1997ARA&A..35..309F,2017hsn..book..195G}, further classified on the basis of their photometric evolution after peak \citep{1979A&A....72..287B,2011MNRAS.412.1522S,2011MNRAS.412.1441L,2017ApJ...837..121G}, distinguishing between transients showing a characteristic {\it plateau} lasting $\simeq100\,\rm{d}$ \citep[see, e.g.,][]{2014ApJ...786...67A} and those showing linear, or almost linear declines after maximum light \citep[see, e.g,][and references therein]{2014MNRAS.445..554F}.
Although this diversity might be solely due to different amounts of H retained at the time of the explosion, different progenitor channels have been proposed for the rare class of SNe IIL \citep[$6-10$\% of all CC SNe;][]{2011MNRAS.412.1522S,2011MNRAS.412.1441L} and the more common Type IIP SNe.
In particular, Type IIL have been proposed to arise from more massive stars, partially depleted of their outer H layers, with larger radii \citep[a few $10^3$\rsun; e.g.,][]{1993A&A...273..106B} with respect to more compact \citep[$<1600$\rsun;][]{2005ApJ...628..973L} and less massive red supergiants (RSGs) \citep[see, e.g.,][]{2010ApJ...714L.254E,2010ApJ...714L.280F,2012MNRAS.424.1372A}, although \citet{2017ApJ...838...28M} showed that RSGs surrounded by a dense CSM can also produce the observables of SNe IIL.
On the other hand, more recently, a few authors proposed Type II SNe to form a heterogeneous class, with their light curves forming a continuum of properties \citep{2014ApJ...786...67A,2015ApJ...799..208S,2016AJ....151...33G,2016ApJ...828..111R,2016MNRAS.459.3939V,2019MNRAS.490.2799D}.

CC SNe interacting with a dense H-rich circumstellar medium (CSM) typically show narrow features (with a {\it full--width--at--half--maximum} -- FWHM of a few $10^2$ up to a few $10^3$\kms) and are therefore labelled as SNe IIn \citep{1990MNRAS.244..269S}.
These are recombination lines emitted by the outer un-shocked CSM, with ionizing photons produced in the underlying shocked regions \citep[see][]{1994ApJ...420..268C}.
Narrow lines, on the other hand, are not the only signature of ongoing ejecta-CSM interaction, as collisions between dense shells are expected to produce boxy, flat topped profiles \citep[see, e.g.,][and references therein]{2011MNRAS.417..261I,2017hsn..book..795J}, and strong signatures of interaction can be deduced from observations in the X-ray and radio domains of transients not showing narrow emission features \citep[see, e.g.,][]{1996ApJ...461..993F}.

\begin{figure}
\begin{center}
\includegraphics[width=\columnwidth]{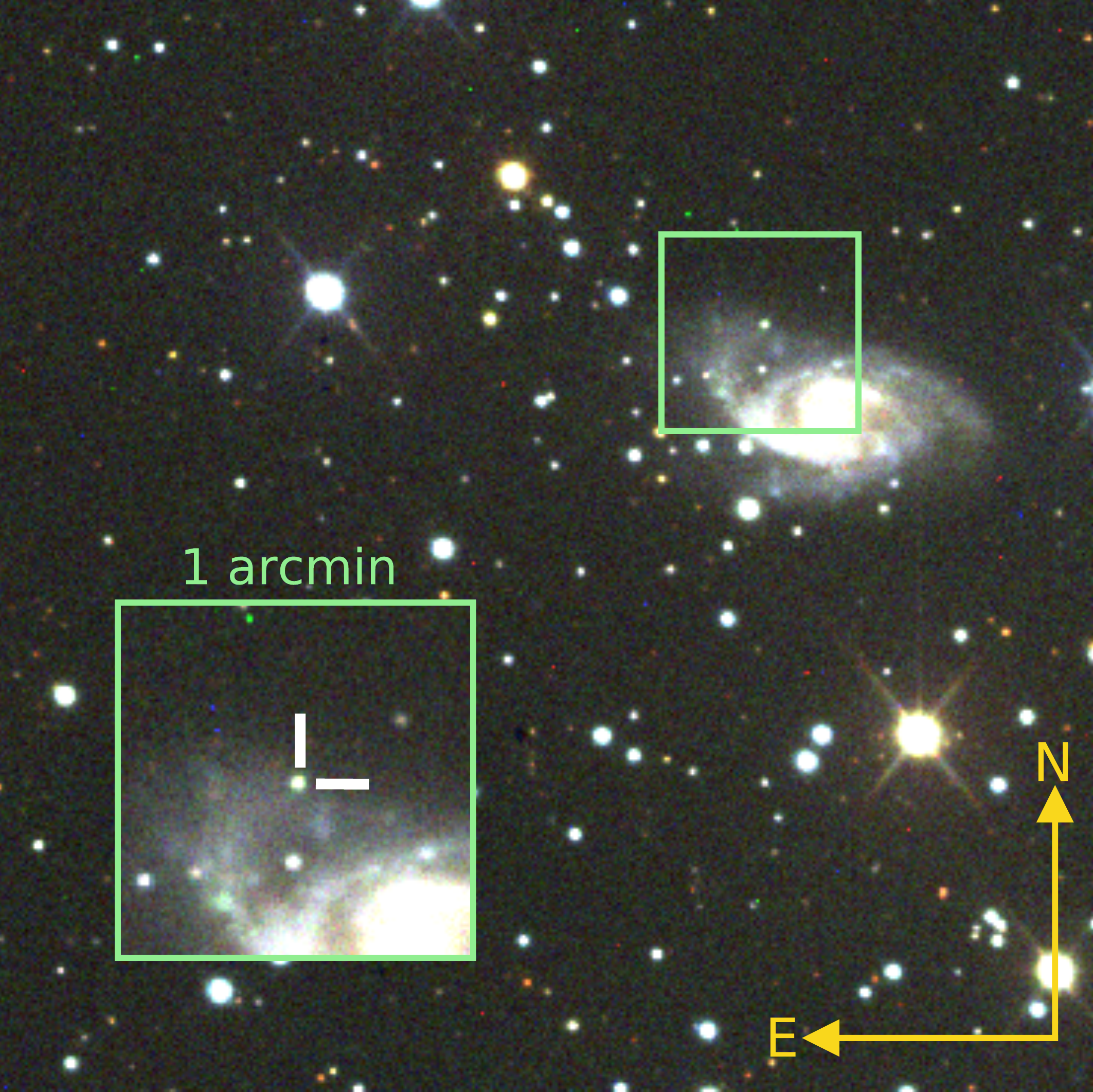}
\caption{Color image of SN~2017ahn and its host galaxy NGC~3318. The image combined $g-$, $r-$ and $i-$band data obtained on 2017 April 16 ($\sim68\,\rm{d}$ days after explosion) with a $1\,\rm{m}$ telescope of the Las Cumbres Observatory network (1m-012, node at the South African Astronomical Observatory -- SAAO, Cape Town, South Africa). SN~2017ahn is the bright source in the middle of the inset. \label{fig:17hFC}}
\end{center}
\end{figure}
While in ``normal" CC SNe the SN shock is expected to break through the stellar photosphere, in stars exploding within a dense medium this typically occurs within the CSM \citep[see, e.g.,][]{2012ApJ...759..108S,2018NatAs...2..808F}, leading to a drastic increase in the time scale of the {\it shock breakout} signal \citep{2011MNRAS.414.1715B}.
While this signal typically fades within seconds to a fraction of an hour after explosion, in SNe interacting with a dense CSM this can be extended up to a time scale of days \citep{2011MNRAS.414.1715B}.
This is the case of Type IIn SNe, where the shock can break through the extended CSM up to a hundred of days after the SN explosion \citep[see, e.g.,][]{2020A&A...635A..39T}, as long as the optical depth of the overlying medium is larger than $\simeq c/v$ (where $v$ is the shock expansion velocity).
After this time, their photometric evolution is mainly shaped by ongoing ejecta-CSM interaction, depending on the efficiency in the conversion of kinetic energy into radiation and the density profiles of the SN ejecta and shocked gas \citep[see][]{1982ApJ...258..790C,2013MNRAS.435.1520M,2014ApJ...797..118F}.
Similarly, their spectroscopic evolution can be dominated by interaction up to many years after the SN explosion, with line profiles shaped by electron scattering \citep[see][]{2018MNRAS.475.1261H} or, occasionally, showing emission components from shocked shells \citep[see, e.g.,][]{2020A&A...638A..92T}.
Interaction can mask the underlying ejecta, preventing us from collecting information about the progenitor stars (typically accessible during the nebular phases) and, occasionally, even explosion mechanisms \citep[see, e.g.,][]{2013ApJS..207....3S}.
This is mainly because of the pseudo photosphere being typically located in the outer un-shocked CSM, produced through stationary winds or eruptive events throughout the evolution of the progenitor star, and only reflecting the composition of the most outer layers of its envelope. 
\begin{figure}
\begin{center}
\includegraphics[width=\columnwidth]{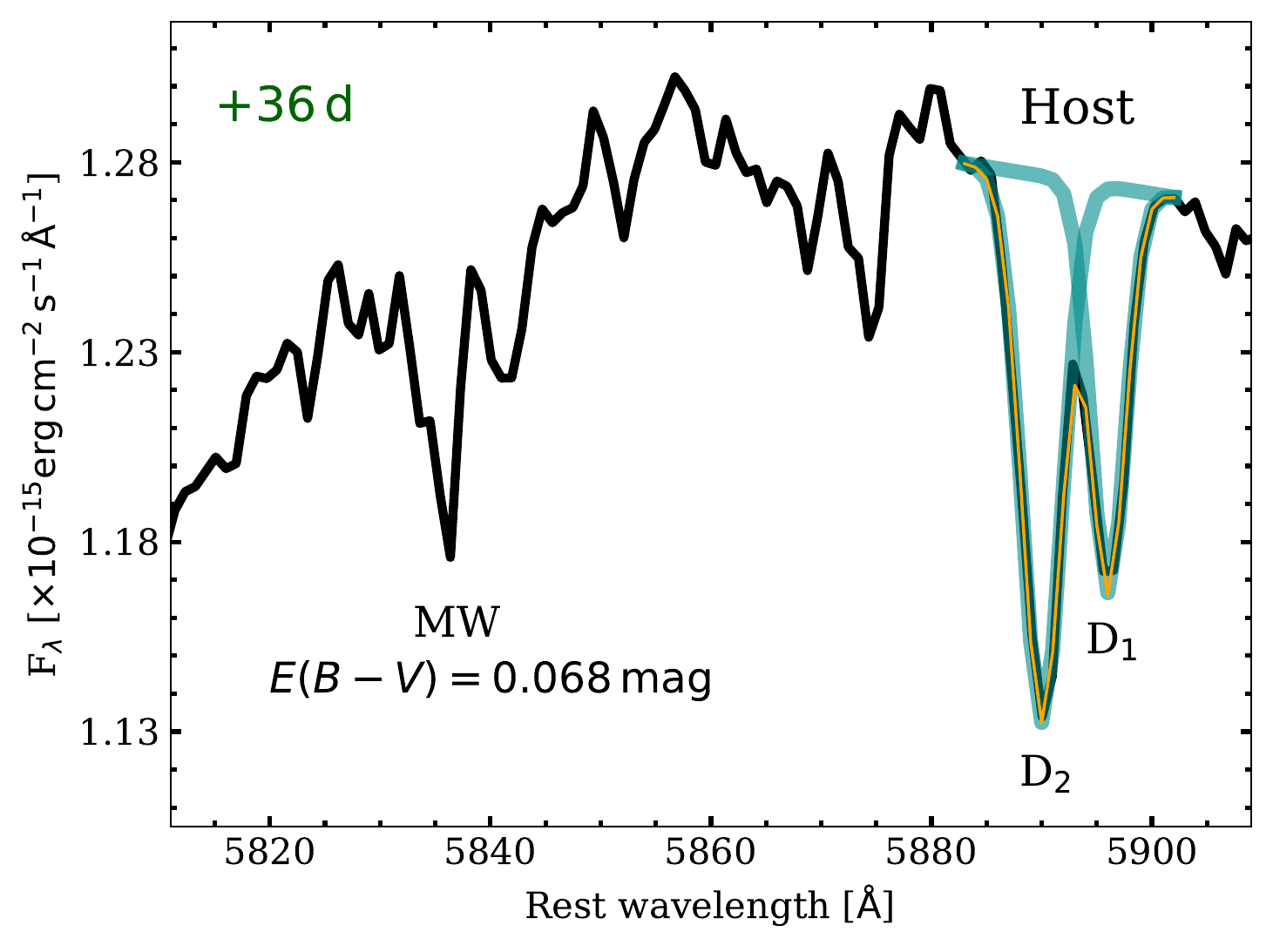}
\caption{Zoom-in of the $5820-5900$\ang~region of the spectrum obtained on 2017 March 15 ($36\,\rm{d}$ after explosion). Galactic \ion{Na}{1D} and host absorption features are clearly visible. A multi-gaussian fit (yellow line) of the marginally resolved host features including the single components (cyan lines). The spectrum is corrected for the recessional velocity of \object. \label{fig:sodium}}
\end{center}
\end{figure}

On the other hand, in CC SNe discovered soon after explosion, narrow features may arise from shells expelled during the very late phases of the evolution of their progenitors, and hence reflect the chemical composition of their outer layers just before explosion (see, e.g., the cases of SNe~1983K; \citealt{1985ApJ...289...52N} 1993J; \citealt{1994AJ....108.1002G,2000AJ....120.1499M}, 1998S; \citealt{2000ApJ...536..239L,2015ApJ...806..213S}, 2006bp \citealt{2007ApJ...666.1093Q}, and the more recent cases of SNe~2013cu; \citealt{2014Natur.509..471G}, 2013fs; \citealt{2017NatPh..13..510Y} and 2016bkv; \citealt{2018ApJ...861...63H}). 
These high ionization (e.g., \ion{He}{2}, \ion{C}{3-IV}, \ion{N}{4-V} and occasionally \ion{O}{4-VI}) features -- sometimes dubbed `flash spectroscopy' features -- rapidly fade and typically disappear after a few days, depending on the physical conditions of the emitting shell, and on the time these regions are overtaken by the rapidly expanding SN ejecta.
The occurrence of CC SNe showing early high-ionization features is expected to be relatively high \citep[$\sim20$\% among those discovered within $5\,\rm{d}$ since explosion;][]{2016ApJ...818....3K} and their numbers will increase with the advent of modern $\sim$1 day cadence SN surveys such as the Palomar Transient Facility \citep[PTF and its continuation iPTF;][]{2009AAS...21346901L,2013ATel.4807....1K}, the All Sky Automated Survey for SNe \citep[ASAS-SN; ][]{Shappee14}, the Asteroid Terrestrial-Impact Last Alert System \citep[ATLAS;][]{Tonry2011,ATLAS2}, the Distance Less Than $40\,\rm{Mpc}$ \citep[DLT40;][]{2018ApJ...853...62T} and the Zwicky Transient Facility \citep[ZTF;][]{2019PASP..131g8001G,2019PASP..131a8002B}. 

\begin{figure}
\begin{center}
\includegraphics[width=\columnwidth]{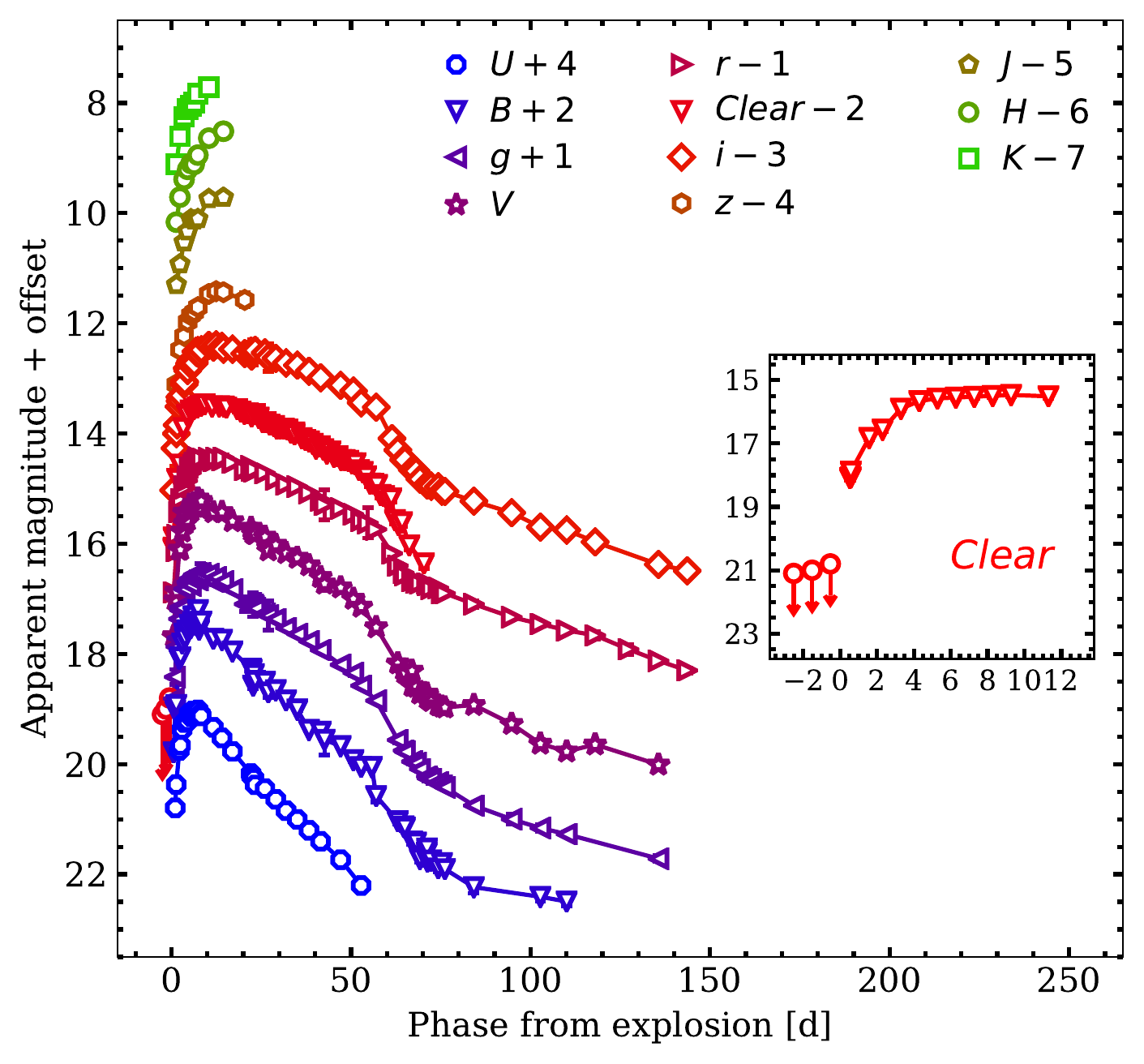}
\caption{Optical and NIR light curves of \object. $U,B,V,Clear,J,H,K$ and $g,r,i,z$ magnitudes were calibrated to the Vega and AB photometric systems, respectively. Magnitudes were not corrected for the foreground Galactic or host extinction. Phases refer to the estimated epoch of the explosion. In the inset, a zoom-in shows the last non-detection limits and the early evolution of the DLT40 data. \label{fig:optNIRphoto}}
\end{center}
\end{figure}
In this context, we present the discovery and the detailed follow-up campaign of the Type II \object.
\object~was discovered on 2017 February 8.29~UT \citep{2017ATel10058....1T} in the nearby galaxy NGC~3318 during the second year of operations of DLT40, specifically looking for nearby SNe within one day from explosion. 
It was given an internal designation of  DLT17h.
First detection and the subsequent confirmation image were both obtained using the $0.41\,\rm{m}$ PROMPT~5 telescope \citep{prompt} located at the Cerro Tololo Inter-American Observatory (CTIO, Cerro Pach\'on, Chile). 
\object~was also observed at radio frequencies on 2017 February 28.6~UT ($\rm{JD}=2457813.1$, $\simeq21\,\rm{d}$ after explosion) with the Australia Telescope Compact Array (ATCA), resulting in non-detection limits of 75 and $40\,\rm{\mu Jy}\,\rm{beam^{-1}}$ at 5.5 and $9.0\,\rm{GHz}$, respectively \citep{2017ATel10147....1R}.
Further details about the DLT40 survey during the time period of this SN discovery are discussed in \citet{Yang17,2018ApJ...853...62T,Yang19}.

In Section~\ref{sec:host} we discuss the local environment of \object, and infer its host extinction, while Section~\ref{sec:obsev} includes details about the  photometric (\ref{sec:lightcurves}) and spectroscopic (\ref{sec:optspectra} and \ref{sec:NIRspectra}) follow-up campaigns of \object.
In Section~\ref{sec:analysis}, we discuss the main observables  in the context of young nearby CC SNe, while the main results of our analysis are summarized in Section~\ref{sec:conclusions}.
Additional information about the facilities used to collect data as well as the reduction steps and tools are reported in the Appendix.

\section{The local environment} \label{sec:host}
\object~is located at $\rm{RA}=10$:37:17.45, $\rm{Dec}=-41$:37:05.27 [J2000], 21\farcs75~E, 33\farcs93~N from the center of its host galaxy, NGC~3318 (see Figure~\ref{fig:17hFC}).
NGC~3318 is a nearby ($D\lesssim38\,\rm{Mpc}$\footnote{\url{https://ned.ipac.caltech.edu/}}) spiral galaxy \citep[SAB(rs)b;][]{1991rc3..book.....D}, which already hosted the Type II SN~2000cl \citep{2000IAUC.7432....1G}, 15\farcs47~W, 42\farcs51~S away from the position of \object~and the Type II SN~2020aze (24\farcs44~W, 23\farcs65~S from \object; Ailawadhi et al. in preparation).
In the following, we will assume a luminosity distance of $33.0\pm6.5\,\rm{Mpc}$ to NGC~3318, corresponding to a distance modulus of $\mu=32.59\pm0.43\,\rm{mag}$ (as derived by \citealt{2014MNRAS.444..527S} using data obtained during the observational campaign {\it Cosmicflows with Spitzer}; \citealt{2012AN....333..436C,2012ApJ...749..174C,2012ApJ...749...78T,2013AJ....146...86T}), placing \object~at a projected distance of $\simeq6.4\,\rm{kpc}$ from the center of NGC~3318.
This value is in agreement with that found by \citet{2015MNRAS.450..317C}, resulting in a ``cosmicflows-3" luminosity distance $D_L=40.76\,\rm{Mpc}$ (assuming $\Omega_M=0.3$, $\Omega_\Lambda=0.7$ and $H_0=70\,\rm{km}\,\rm{s^{-1}}\,\rm{Mpc^{-1}}$), corresponding to a distance modulus $\mu=33.05\pm\,\rm{mag}$\footnote{\url{https://cosmicflows.iap.fr}}, as well as that inferred using The Extragalactic Distance Database (EDD\footnote{\url{http://edd.ifa.hawaii.edu/CF3calculator/}}; see \citealt{2020AJ....159...67K}) based on the linear density field model of \citet[][$\mu=32.73\pm0.21\,\rm{mag}$.]{2019MNRAS.488.5438G}.

For the foreground Galactic extinction we adopt the values reported by \citet{2011ApJ...737..103S}, corresponding to $E(B-V)=0.068\,\rm{mag}$, while to estimate the local extinction, we compared results obtained using different methods.

A first estimate was obtained using data collected with the VLT/Multi Unit Spectroscopic Explorer \citep[MUSE;][]{2014Msngr.157...13B} integral field spectrograph on 2015 May 17\footnote{ESO Programme ID 095.D-0172.}, as part of a survey of nearby SN explosion sites \citep[see][for details]{2018A&A...613A..35K}, which serendipitously included the site of \object~in its field of view.
Line-of-sight extinction was estimated from the observed Balmer decrement (\ha/\hb, after correcting the spectrum for the Galactic extinction) of a nearby \ion{H}{2} region, assuming an intrinsic flux ratio of 2.86 \citep[Case B recombination;][]{2006agna.book.....O} and a standard extinction law with $R_V=3.1$ \citep{1989ApJ...345..245C}, yielding an additional contribution of $E(B-V)=0.09\pm0.06\,\rm{mag}$ from the local environment.

On the other hand, a rapid inspection of the $5800-6000$\ang~region in our spectrum obtained on 2017 March 15 ($36\,\rm{d}$ after explosion, see Section~\ref{sec:obsev}) reveals host \ion{Na}{1D} features much stronger than the those due to the Galactic extinction, with a total equivalent width $\rm{EW_{host}}\simeq2.6\times\rm{EW_{MW}}$, as measured from the overall profile including both lines (see Figure~\ref{fig:sodium}).
This ratio would suggest an additional contribution of $\rm{E(B-V)\simeq0.18\,\rm{mag}}$ (assuming a standard extinction law) of the local environment to the total reddening. 
Since at this phase the feature is marginally resolved, we fit its profile using a combination of two gaussian centered at the position of the D2 and D1 \ion{Na}{1D} lines (see Figure~\ref{fig:sodium}).
The fit gives parameters for both absorption features, including their EWs, resulting in $\simeq0.53$ and $\simeq0.33$\ang~for D2 and D1, respectively.
These are both within the linearity range of the relation between the sodium EW and dust extinction derived by \citet{2012MNRAS.426.1465P} (see their Figure~9).
Taking a weighted average between the quantities inferred using their Equations~7 and 8, these correspond to a contribution of $E(B-V)=0.196\pm0.054\,\rm{mag}$ to the total reddening.
A similar result ($E(B-V)\simeq0.12\,\rm{mag}$) is obtained using the relations found by \citet[][see also \citealt{2009ApJ...693..207B}]{2003fthp.conf..200T}.

In order to avoid possible projection effects (\object~lies at a projected distance of $\simeq160\,\rm{pc}$ from the center of the \ion{H}{2} region, assuming a distance of $33\,\rm{Mpc}$), we therefore favor $E(B-V)=0.196\,\rm{mag}$ as the contribution of the local environment to the total color excess in the direction of \object.
This value has to be summed to the Galactic reddening in the direction of \object, resulting in a total extinction $E(B-V)=0.264\pm0.054\,\rm{mag}$, although we cannot rule out a lower extcintion value (see Section~\ref{sec:CMFGEN}).
In addition, as shown by \citet{2013ApJ...779...38P}, these relations largely underestimate errors on the derived values, and hence the uncertainty on the reddening is probably larger.

\section{Evolution of the main observables} \label{sec:obsev}
\subsection{Light curves} \label{sec:lightcurves}
Optical and NIR light curves are shown in Figure~\ref{fig:optNIRphoto}, while the corresponding magnitudes and a description of the facilities used and reduction steps are reported in Appendix~\ref{sec:photoredu}.

The explosion epoch was estimated from the unfiltered light curve obtained with PROMPT5 (see the inset in Figure~\ref{fig:optNIRphoto}), taking the mid-point between the discovery (2017 February 8.29~UT; $\rm{JD}=2457792.79$) and the last non-detection limit ($m>20.8\,\rm{mag}$ on 2017 February 7.23~UT; $\rm{JD}=2457791.73$).
In the following, we will therefore assume 2017 February 7.76~UT ($\rm{JD}=2457792.26\pm0.5$) as the explosion epoch for \object, and refer phases to this date.
\begin{figure}
\begin{center}
\includegraphics[width=\columnwidth]{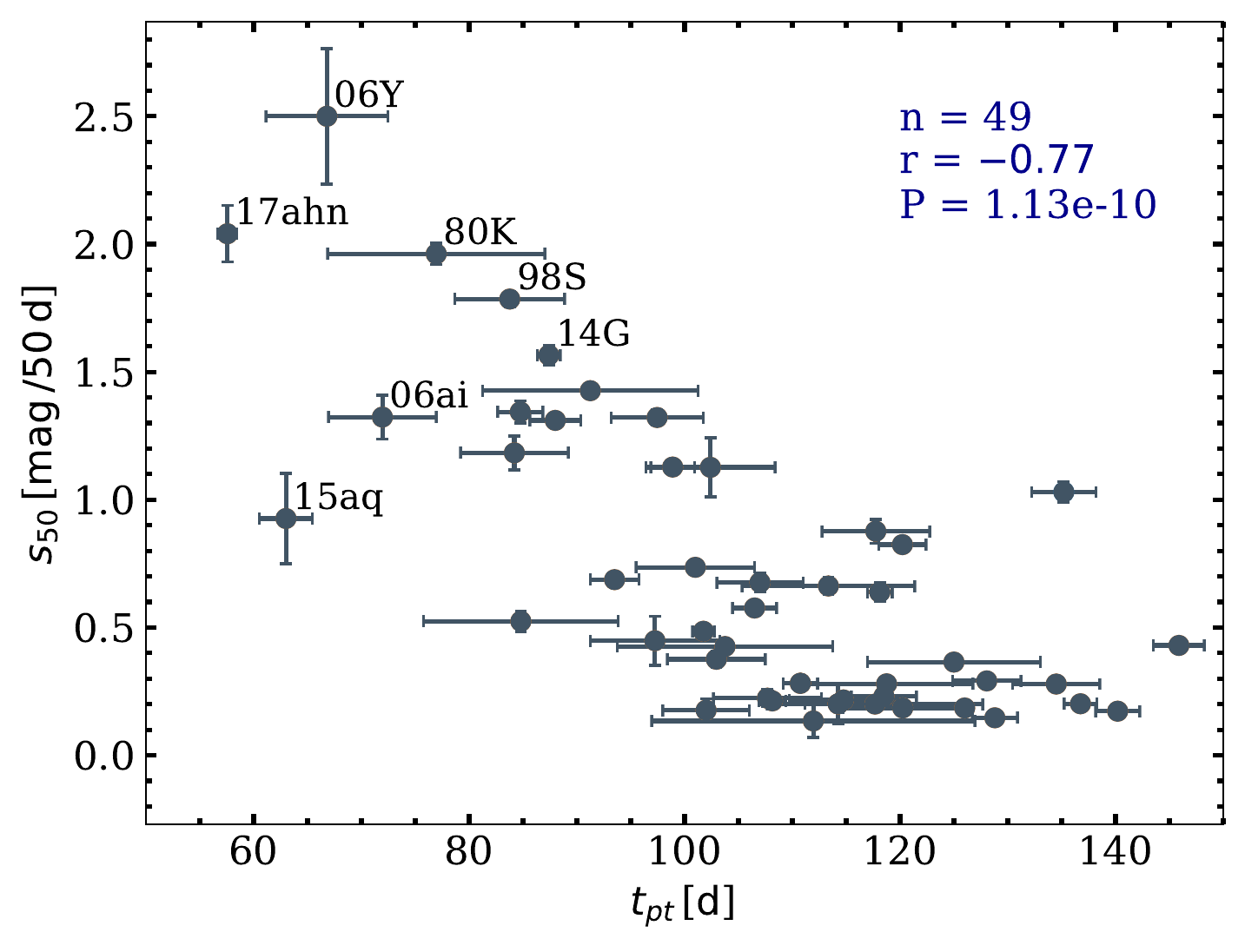}
\caption{$V-$band decline rate measured at $+50\,\rm{d}$ ($s_{50}$) vs. $t_{pt}$ (see the main text) for the sample of Type II SNe of \citet{2016MNRAS.459.3939V}, including \object~and SN~1998S. n is the number of objects in the sample, r the Pearson $s-$correlation coefficient and P the probability of getting a correlation by chance. \label{fig:s50tpt}}
\end{center}
\end{figure}

Optical light curves show a relatively fast rise to maximum \citep[see, e.g.,][]{2014ApJ...786...67A,2015A&A...582A...3G,2015MNRAS.451.2212G,2016MNRAS.459.3939V} in all bands, with an average rate of $\sim0.8\,\rm{mag}\,\rm{d^{-1}}$.
Fitting high order polynomials, we infer $t_{rise}=t_{max}-t_{expl}$ ranging from $6.38\pm0.66\,\rm{d}$ in $U$ to $8.03\pm0.71\,\rm{d}$ in $V$, with a similar behavior in $griz$ ($8.54\pm2.24\,\rm{d}\leq t_{rise}\leq14.04\pm3.99\,\rm{d}$).
Errors were estimated performing Monte Carlo simulations, randomly shifting the photometric data within their uncertainties, including that on the estimated epoch of the explosion.

After peak, light curves settle on a short plateau, lasting $\simeq50\,\rm{d}$, more pronounced at redder wavelengths, except for the $U-$band, declining linearly (with a rate of $0.069\pm0.001\,\rm{mag}\,\rm{d^{-1}}$).
At $t\gtrsim+70\,\rm{d}$, the optical light curves settle on a tail, with a slower decline at an average rate of $\simeq0.02\,\rm{mag}\,\rm{d^{-1}}$, with the exception of the $B-$band light curve, showing a late decline of $\simeq0.01\,\rm{mag}\,\rm{d^{-1}}$. 
This suggests a luminosity evolution faster than that predicted by the radioactive $^{56}\rm{Co}$ decay (see Section~\ref{sec:analysis}). 
Following \citet{2016MNRAS.459.3939V} we infer a decline rate $s_{50,V}=2.04\pm0.11\,\rm{mag}\,50\,\rm{d^{-1}}$ and a mid-point between the end of the plateau and the onset of radioactive tail $t_{pt,V}=57.56\pm0.88\,\rm{d}$.
These are similar to other fast declining Type II SNe such as SNe~1980K, 2006Y \citep{2014ApJ...786...67A}, 2014G \citep{2016MNRAS.462..137T} and 1998S \citep{2000MNRAS.318.1093F} (see Figure~\ref{fig:s50tpt}).
\begin{figure}
\begin{center}
\includegraphics[width=\columnwidth]{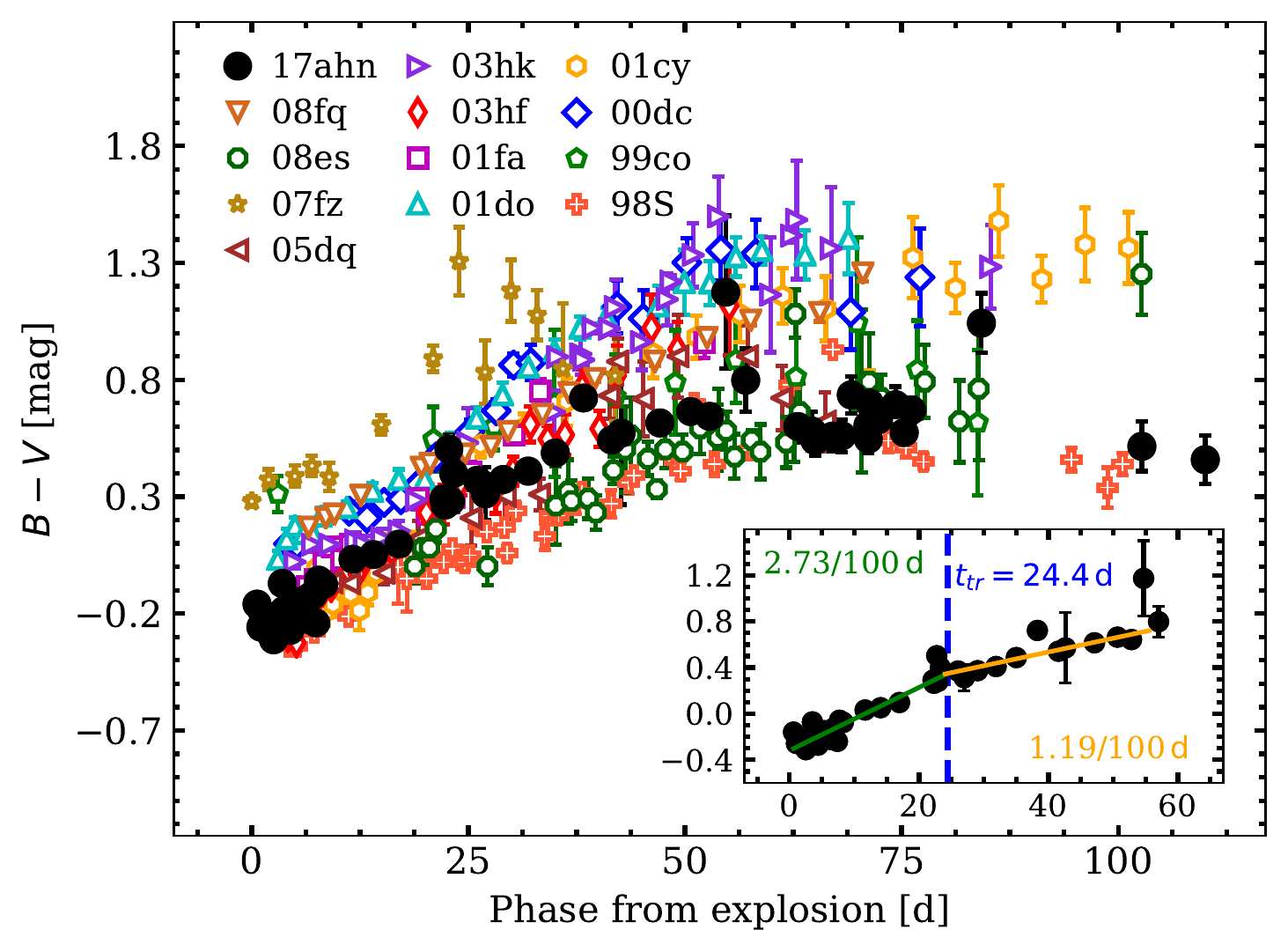}
\caption{$B-V$ color curve of \object~compared to the color evolution of the sample of \citet{2014MNRAS.445..554F}. The $B-V$ color evolution of SN~1998S is also included for comparison. Magnitudes were corrected for the total (Galactic+host) extinction. \label{fig:colorIILs}}
\end{center}
\end{figure}

In Figure~\ref{fig:colorIILs}, we compare the $B-V$ colors of \object~to those of the sample of similarly fast declining Type II SNe of \citet{2014MNRAS.445..554F}.
The resulting evolution is consistent with the bluer end of the distribution, corresponding to colors similar to those shown by SNe~1999co and 1998S, suggesting a higher temperature for the pseudo-photosphere, as will be discussed in the following sections.
Following \citet{1994A&A...282..731P}, we identify two `regimes' with different slopes in the $B-V$ color evolution.
Fitting linear relations, we find an initial evolution towards redder color with a slope of $2.73\pm0.06\,\rm{mag}\,100\,\rm{d^{-1}}$, followed by a slower evolution with a rate of $1.19\pm0.09\,\rm{mag}\,100\,\rm{d^{-1}}$ with a break occurring at $t_{tr}\simeq+24.4\,\rm{d}$.
These slopes are both steeper, with a break in the color evolution of \object~occurring slightly earlier than the median values found for the sample of Type II SN of \citet[][$s_{1,(B-V)}\simeq2.63$ and $s_{2,(B-V)}\simeq0.77\,\rm{mag}\,100\,\rm{d^{-1}}$, respectively, with $t_{tr}\simeq37.7\,\rm{d}$]{2018MNRAS.476.4592D}.
At $t\gtrsim+75\,\rm{d}$ we note a further flattening, with $B-V$ colors remaining roughly constant throughout the remaining $\simeq25\,\rm{d}$ of photometric coverage.

Assuming the distance modulus and total reddening reported in Section~\ref{sec:host}, we infer absolute peak magnitudes ranging from $M_z=-17.81\pm0.29\,\rm{mag}$ to $M_U=-19.08\pm0.29$ (with $M_V=-18.44\pm0.29$), where the errors are dominated by the uncertainty on the distance modulus and extinction (see Section~\ref{sec:host}).

\begin{figure}
\begin{center}
\includegraphics[width=\columnwidth]{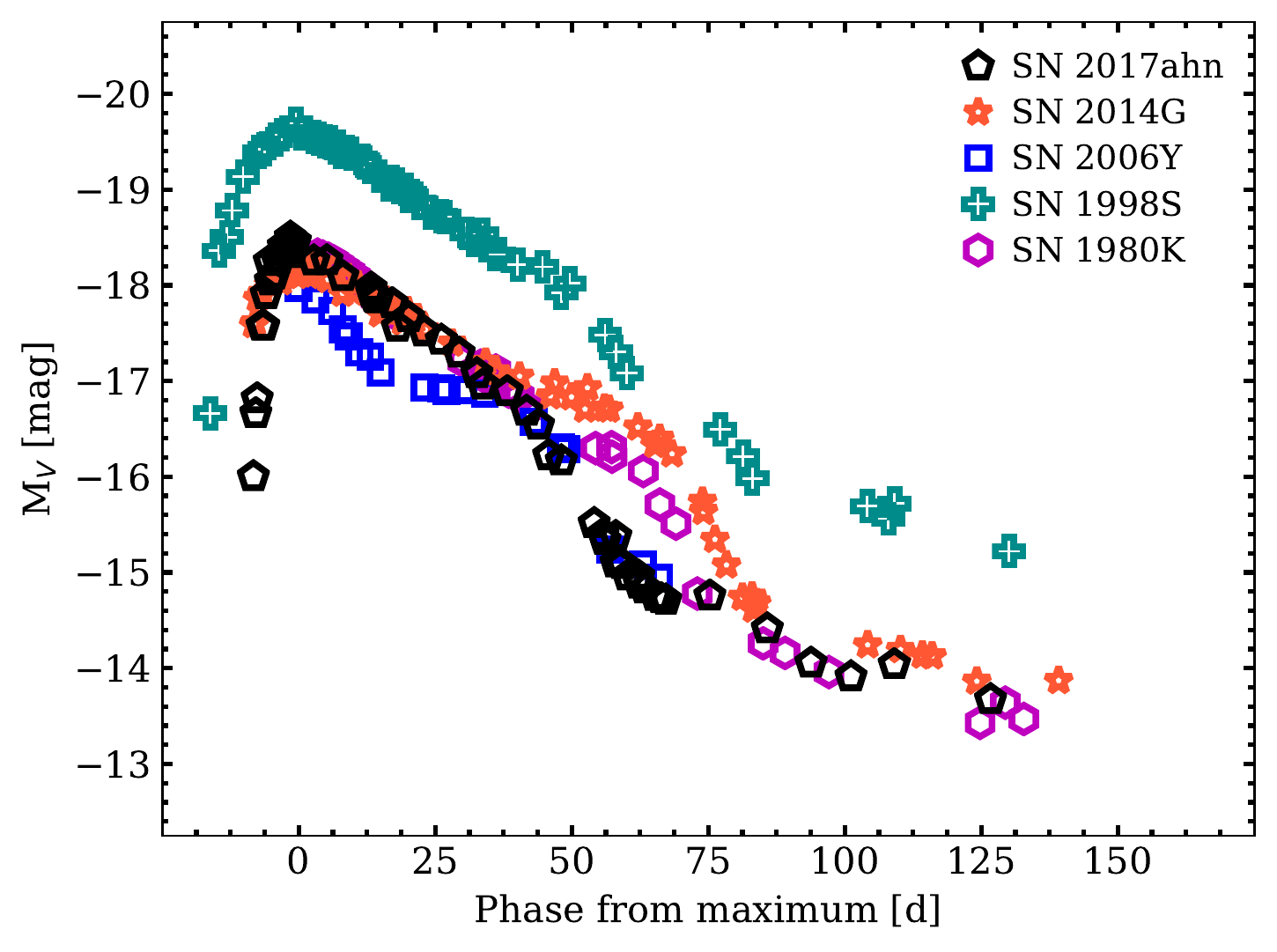}
\caption{Absolute $V-$band light curve of \object~compared to those of other transients showing a similar photometric evolution (see the main text). \label{fig:amcomp}}
\end{center}
\end{figure}

A comparison with other transients showing similar photometric properties, based on the results reported above (see Figure~\ref{fig:amcomp}), reveals an evolution of the absolute $V-$band magnitude similar to that of SN~1998S \citep{2000MNRAS.318.1093F,2000A&AS..144..219L}, which, despite the systematically brighter magnitudes and a slightly different rise time, shows the same relatively short plateau, with a comparable fast decline at $t\gtrsim+70\,\rm{d}$ ($0.002\pm0.03\,\rm{mag}\,\rm{d^{-1}}$).

Given the similarities in the photometric evolution of \object~and SN~1998S and the limited coverage of our NIR light curves, we cannot rule-out the presence of an IR excess similar to those typically observed in long-lasting Type IIn SNe \citep[see, e.g.,][]{2002ApJ...575.1007G,2011ApJ...741....7F} and in SN~1998S itself \citep[see][]{2004MNRAS.352..457P}.
These features are often explained either as `light echoes' by pre-existing dust \citep[see, e.g.,][]{2020A&A...635A..39T} or newly formed dust in the post-shocked regions \citep[e.g., ][]{2012AJ....143...17S}.
On the other hand, the absence of colder components in the NIR spectral continuum of \object~seems to suggest a lack of a clear IR excess at least until $+65\,\rm{d}$ (see Section~\ref{sec:analysis}).

The field of \object~was also observed using the Ultraviolet/Optical Telescope (UVOT) on board the Neil Gehrels $Swift$ Observatory \citep{Gehrels04}, obtaining 8 epochs covering the early photometric evolution of \object~(up to $+15\,\rm{d}$).
The resulting light curves are shown in Figure~\ref{fig:UVOTcurves}, while details about the reduction steps are given in Appendix~\ref{sec:photoredu}.
The light curves show a short and steep rise, more pronounced at bluer wavelengths, where we measure an average increase in luminosity of $\simeq0.5\,\rm{mag}\,\rm{d^{-1}}$ in $W2,M2$ and $W1$, respectively, peaking at $t_{max}\simeq+3.5\,\rm{d}$ ($W2,M2$) and $+5.3\,\rm{d}$ ($W1$).
Rising UV light curves can either be interpreted by the simultaneous fast expansion and cooling of an extremely hot pseudo-photosphere ($T>1.5\times10^{4}\,\rm{K}$) or by an intrinsic increase of the UV luminosity and temperatures, like those expected in the shocked regions of interacting transients.
\begin{figure}
\begin{center}
\includegraphics[width=\columnwidth]{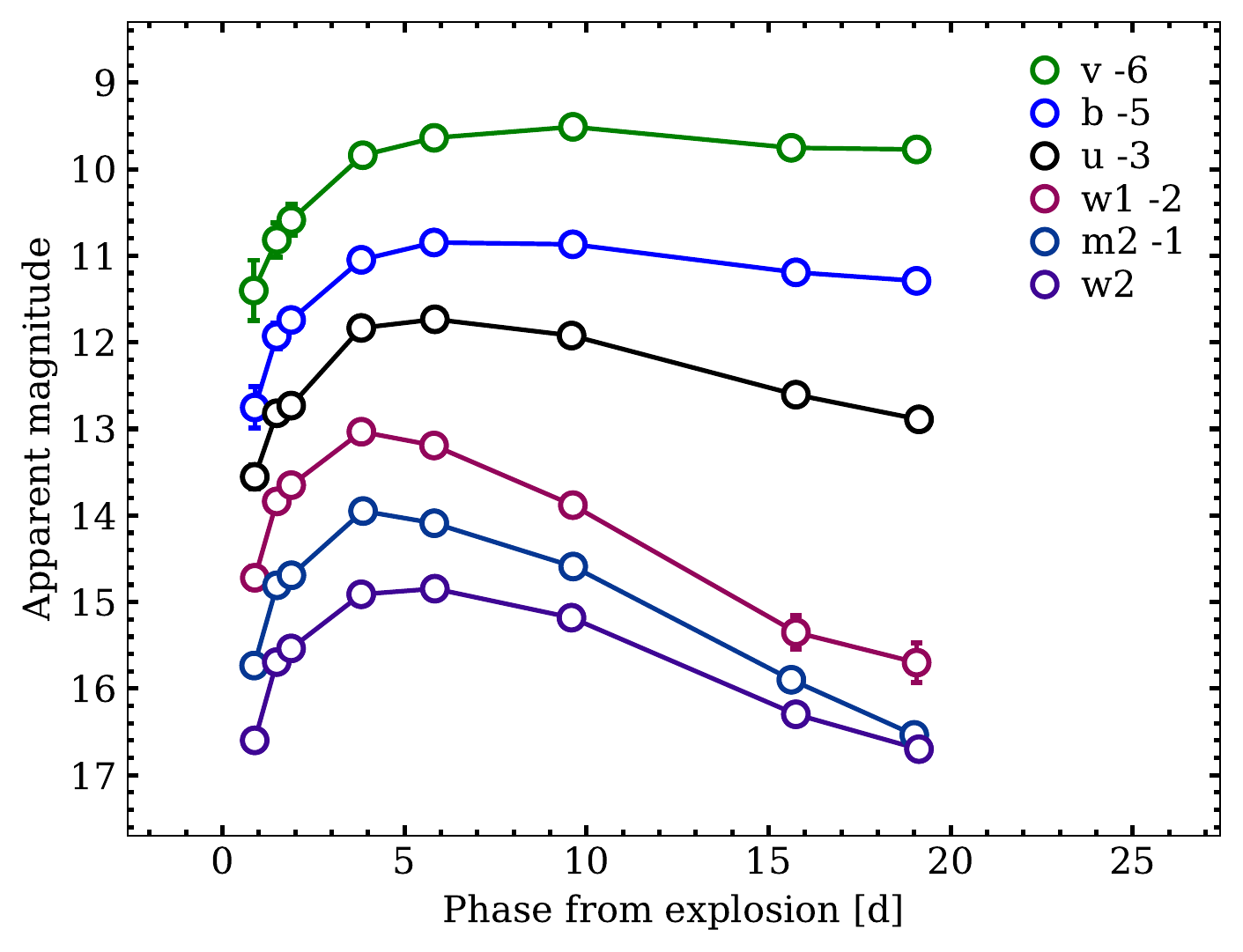}
\caption{{\it Swift} UVOT light curves of \object. Magnitudes were calibrated to the Vega photometric system and have not been corrected for extinction.\label{fig:UVOTcurves}}
\end{center}
\end{figure}
\begin{figure*}
\begin{center}
\includegraphics[width=\linewidth]{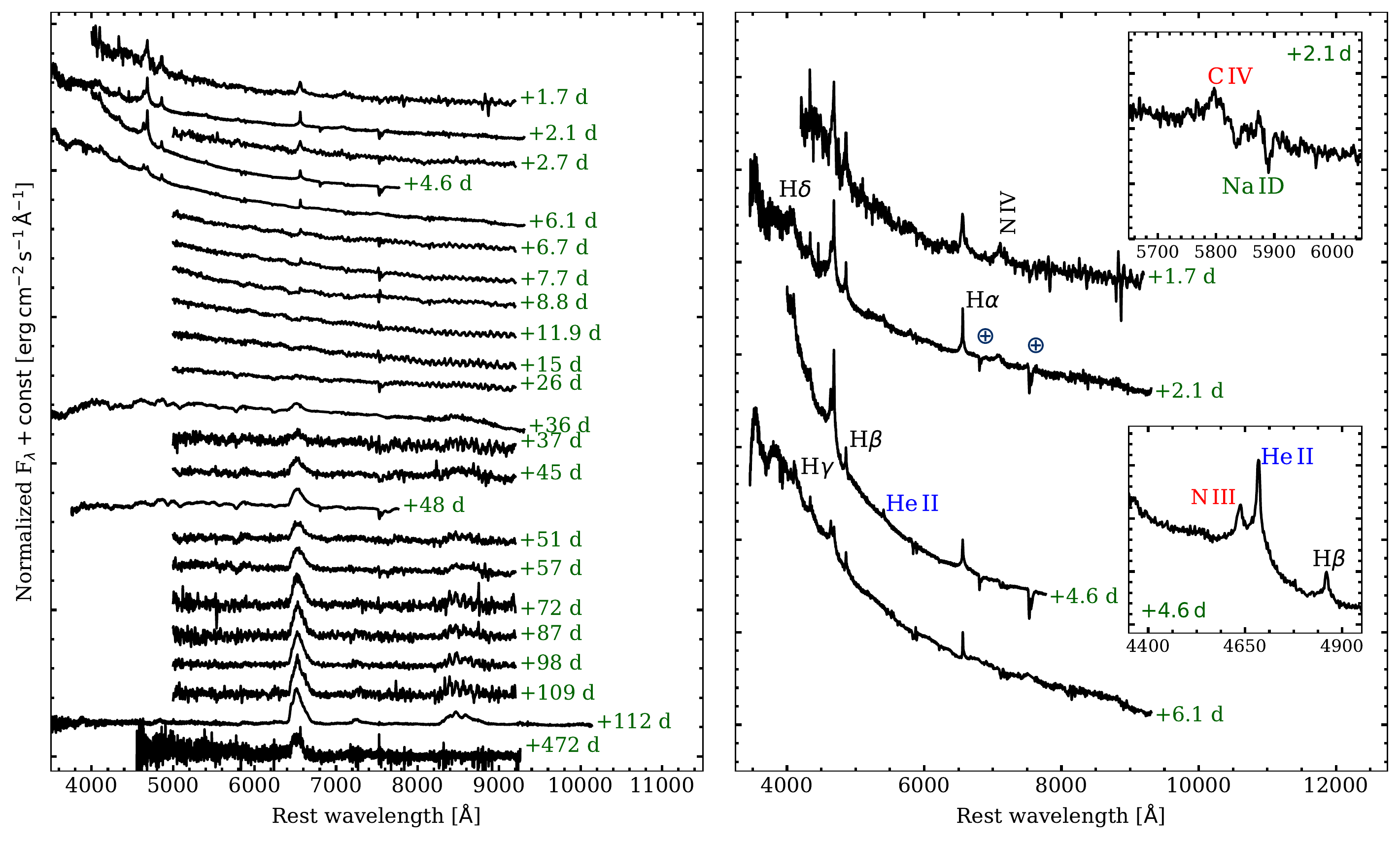}
\caption{{\bf Left:} Optical spectra of \object. Spectra were corrected for the total reddening along the line of sight. {\bf Right:} Early spectroscopic evolution of \object. The spectrum at $+1.7\,\rm{d}$ is re-binned to a third of its original resolution to facilitate the visualization of the high-ionization features (see the main text). The most prominent features are identified. $\oplus$ symbols mark the position of the main telluric absorption features. Insets show zoom-in regions around \ion{Na}{1D} ({\bf upper} inset, including \ion{C}{4} $\lambda5801$) at $+2.1\,\rm{d}$ and \ion{He}{2} $\lambda4686$ ({\bf bottom} inset, including \ion{N}{3}) at $+4.6\,\rm{d}$. Spectra used to create this figure are available as Data behind the Figure. \label{fig:optSpectra}}
\end{center}
\end{figure*}
An upper limit in the X-ray counts was determined using aperture photometry through the HEAsoft packages (\textit{xselect}; \citealt{blackburn_1995} and \textit{xspec}; \citealt{arnaud_1996}).
The background was selected as a region outside the host galaxy with no known X-ray sources within, and measured fluxes were converted from counts per second to luminosities using PIMMS \citep{mukai_1993}.
No significant detections were found over an 18\arcsec aperture integrating over all available SWIFT/XRT epochs.
This resulted in a limiting count rate of $0.904\times10^{-3}\,\rm{counts}\,\rm{s^{-1}}$, which, assuming a power-law model with a photon index of two and a Galactic absorption of $5.89\times10^{20}\,\rm{cm^{-2}}$ \citep{kalberla_2005}, corresponds to an unabsorbed flux of $3.25\times10^{-14}\,\rm{erg}\,\rm{cm^{-2}}\,\rm{s^{-1}}$ ($0.3-10\,\rm{keV}$) and a luminosity of $6.63\times10^{39}\,\rm{erg}\,\rm{s^{-1}}$ at $33\,\rm{Mpc}$.

\subsection{Optical spectra} \label{sec:optspectra}
Optical spectroscopy of \object~was triggered soon after its discovery, with the first spectrum taken $+1.7\,\rm{d}$ after explosion (although an earlier NIR spectrum was obtained at $+1.4\,\rm{d}$, see Section~\ref{sec:NIRspectra}).
Final spectra are shown in Figure~\ref{fig:optSpectra}, while facilities used and reduction steps are described in Appendix~\ref{sec:specredu}.

Early spectra show a very blue continuum with prominent narrow lines in emission, along with strong \ion{Na}{1D} lines at the redshift of the host galaxy.
The \ion{Na}{1D} features are usually related to a non-negligible reddening along the line of sight of the transient \citep[see, e.g.,][]{2012MNRAS.426.1465P}, and their EWs ($0.53$ and 0.33\ang~for D2 and D1, respectively), suggest a moderately extinguished environment for \object~(see Section~\ref{sec:host}). 
At $+1.7\,\rm{d}$ and $+2.1\,\rm{d}$, the spectra show a number of narrow Balmer emission lines (\ha~to \hd), along with \ion{He}{2} $\lambda5411$, \ion{C}{4} $\lambda5801$, \ion{N}{3} and \ion{N}{4} features.
\ion{C}{4} and \ion{N}{4} features rapidly fade below the detection limit and are already not visible at $+4.6\,\rm{d}$, when, on the other hand, \ion{He}{2} $\lambda5411$ clearly emerges from the spectral continuum.
High ionization features (i.e., \ion{N}{3,IV}, \ion{C}{4} and \ion{He}{2}) become progressively fainter with time, although the signal--to--noise (S/N) ratios and resolutions of our spectra do not allow us to rule out the presence of these lines at later phases.
At $+6.1\,\rm{d}$ we also notice the appearance of a narrow \ion{He}{1} $\lambda5875$ feature, not visible at later epochs.
\begin{figure*}
\begin{center}
\includegraphics[width=\linewidth]{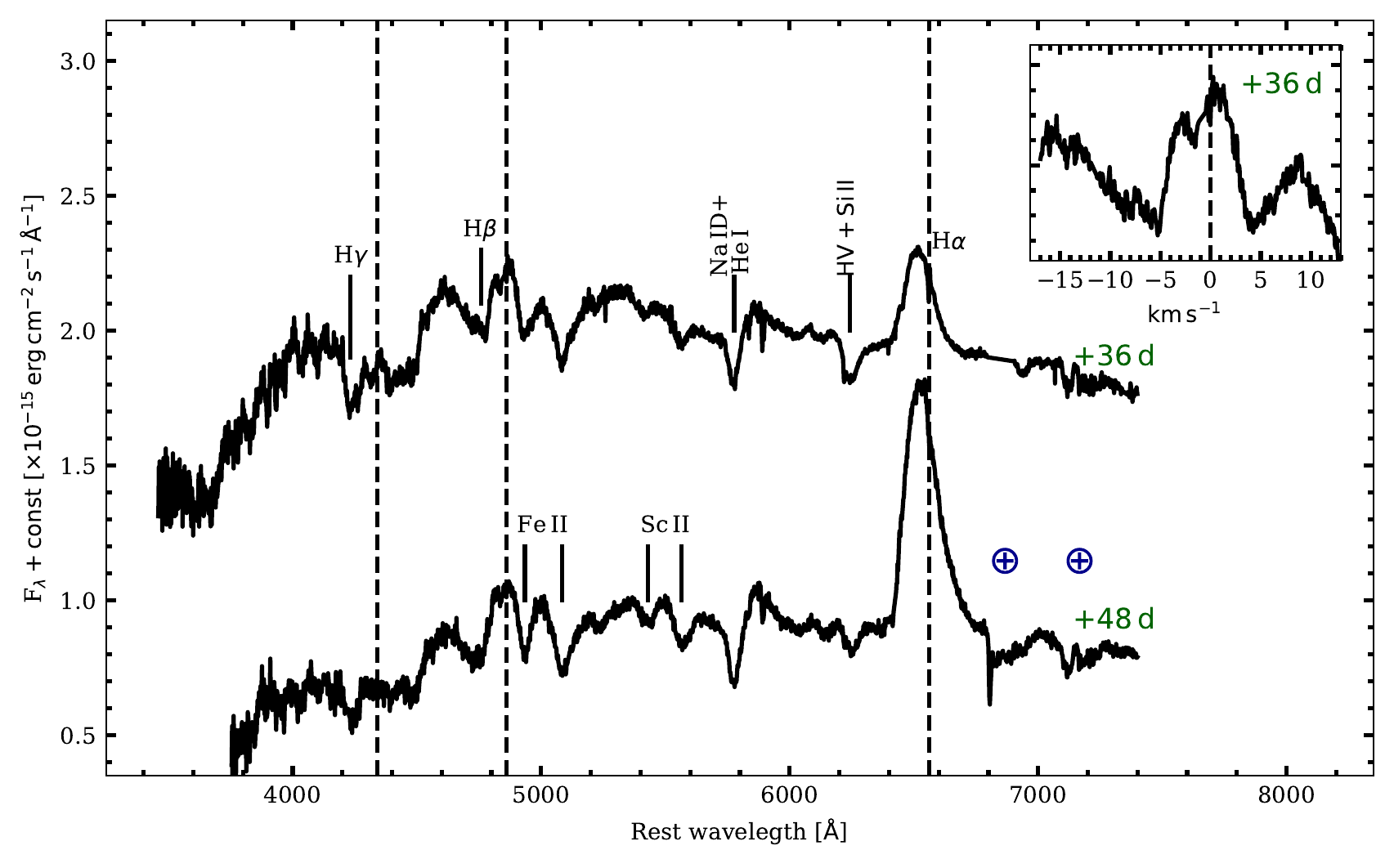}
\caption{Identification of the main spectral features at $+36$ and $+48\,\rm{d}$. Dashed lines correspond to the rest wavelengths of the main Balmer lines, revealing the blue-shift of the emission peaks typical of SNe II \citep{2014MNRAS.441..671A}. $\oplus$ symbols mark the position of the residual telluric absorption features. A possible hydrogen high velocity (HV) feature is also labelled (see the discussion in the main text). In the inset, a zoom-in of the \hb~region in the velocity space, showing the complex structure of the spectral region, with possible multiple absorption features centered at $\simeq-5000$ and $\simeq-10000$\kms~with respect to \hb~rest wavelength. \label{fig:broadID}}
\end{center}
\end{figure*}

The total integrated luminosity (after removing the contribution of the spectral continuum and assuming the reddening reported in Section~\ref{sec:host}) of the \ion{N}{3}+\ion{He}{2} feature shows an initial increase from $\simeq1.4\times10^{40}$\ergs~to $1.6\times10^{40}$\ergs~during the first $+4.6\,\rm{d}$, suggesting an increase in the production of ionizing photons in the underlying regions.
We note the same evolution in the integrated luminosity of \ha, showing an increase of $\Delta L\simeq1.2\times10^{38}$\ergs~over the same period, with a \ha/\hb~ratio evolving from $\simeq0.4$ (at $+2.1\,\rm{d}$) to $\simeq1.9$ (at $+4.6\,\rm{d}$).
At $+6.1\,\rm{d}$ the spectral shape shows the first significant signatures of evolution, with a drastic decrease in the integrated luminosity of the \ion{N}{3}+\ion{He}{2} feature ($\simeq6.4\times10^{39}$\ergs) and the Balmer emission lines (e.g., $L_{\rm{H\alpha}}\simeq1.6\times10^{39}$\ergs, with a \ha/\hb~ratio of $\simeq1.7$).
We also note the appearance of broad and boxy absorption profiles in the blue part of the Balmer lines, with blue wings extending up to $\simeq10^4$\kms, corresponding to expansion velocities of the H-rich shell of $\simeq6500$\kms~(as derived from the minimum inferred through a Gaussian fit to the line profile).

\begin{figure}
\includegraphics[width=\columnwidth]{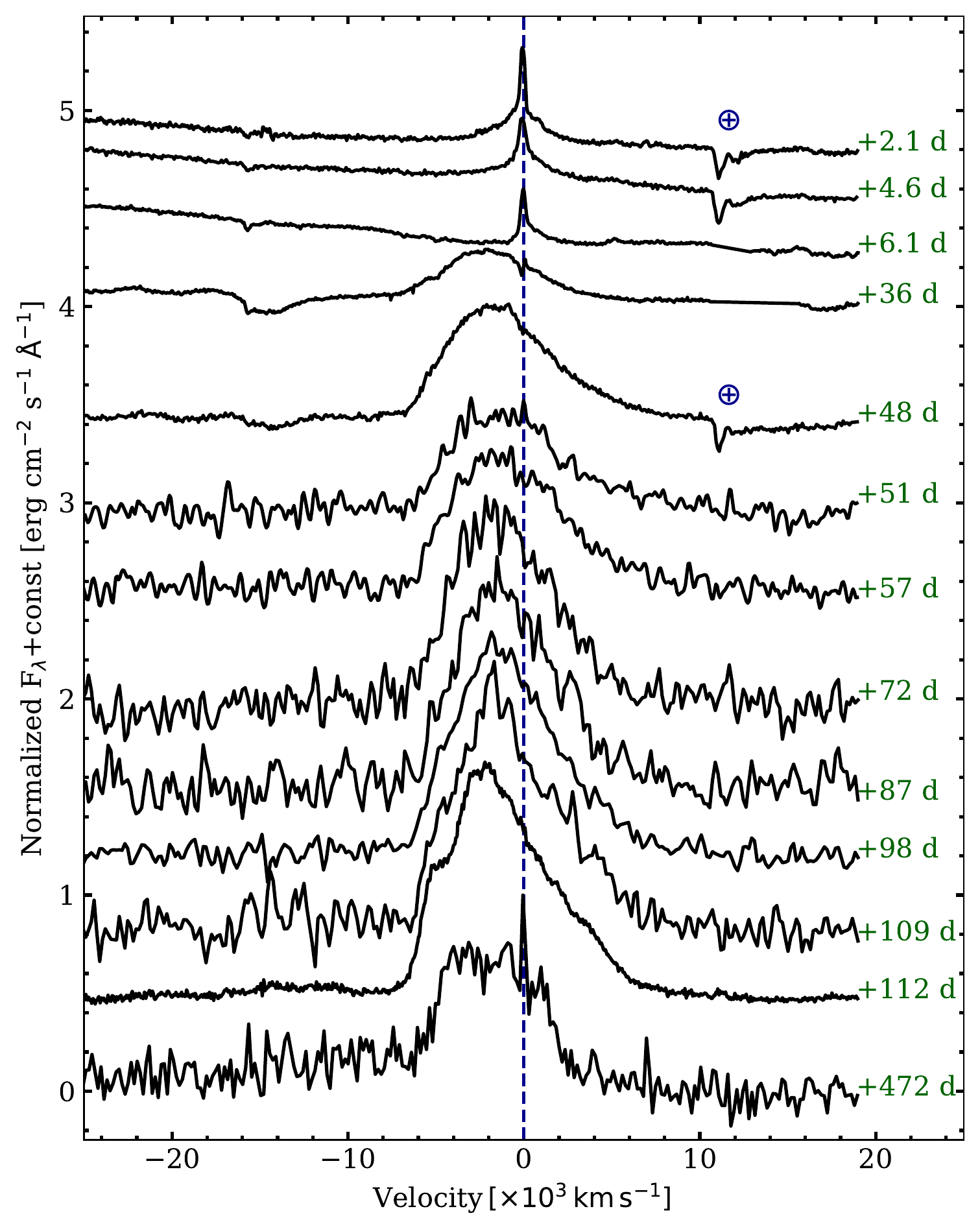}
\caption{Evolution of the profile of \ha~over the first $472\,\rm{d}$ of evolution of \object. Velocities were computed with respect to \ha~rest wavelength. $\oplus$ symbols mark the position of the main telluric absorption features, if visible. The spectrum at $+472\,\rm{d}$ has been degraded to 1/3 of its resolution to facilitate the comparison. \label{fig:halpha}}
\end{figure}

At later epochs, spectra show a further drastic change, with broad P Cygni profiles becoming progressively stronger (see Figure~\ref{fig:broadID}) and the spectral evolution resembling that of a typical Type II SN \citep[see, e.g.,][]{2017ApJ...850...90G}.
At $+36\,\rm{d}$ we clearly see broad \ion{Fe}{2} (multiplet 42) and \ion{He}{1}/\ion{Na}{1D} features. 
At the same phases we also identify \ion{Sc}{2} and a first hint of the presence of the NIR \ion{Ca}{2} triplet, which remains relatively faint throughout the rest of the spectroscopic evolution (see Figure~\ref{fig:optSpectra}), except for the $+112\,\rm{d}$ spectrum, characterized by a significantly higher S/N at the corresponding wavelengths.
At $+112\,\rm{d}$, we also notice broad forbidden \ion{[O}{1]} $\lambda\lambda6300,6364$ and \ion{[Ca}{2]} lines, typical of the nebular phases of Type II SNe, with an integrated luminosity of $\simeq6.4\times10^{38}$\ergs~and $\simeq2.3\times10^{39}$\ergs, respectively.
At $+472\,\rm{d}$ the spectrum only shows a boxy, flat-topped \ha~emission (see Figure~\ref{fig:halpha}), although the S/N does not allow us to rule out the presence of other nebular features at this epoch.
Similarly, the S/N of the spectra at $+51\,\rm{d}\le t\le+109\,\rm{d}$ does not allow us to rule out the appearance of a boxy profile at earlier phases, which would reveal the presence of different CSM layers.

Measuring the positions of the absorption minima at $+36\,\rm{d}$, we infer expansion velocities of $\simeq5000$\kms~from both \hb~and \ion{Fe}{2} $\lambda5169$, the latter usually considered a good proxy of the SN photospheric velocity \citep[see, e.g.,][]{2005A&A...439..671D,2012MNRAS.419.2783T}, although the complex structure of the spectral region around \hb~might suggest higher expansion velocities for the H-rich shell; see the inset in Figure~\ref{fig:broadID}.
We infer similar values from \ion{Sc}{2} lines, while we do not notice a significant evolution in the expansion velocities inferred at $+36$ and $+48\,\rm{d}$.

We obtained a rough estimate of the temperature evolution of the pseudo-photosphere fitting a black-body (BB) function to the spectral continuum, resulting in $T=10000\pm900\,\rm{K}$ in both the $+1.7$ and $+2.1\,\rm{d}$ spectra.
At $+4.6\,\rm{d}$ we note a drastic increase in the temperature ($\simeq31000\pm5000\,\rm{K}$), also reflected by the evolution of the spectral energy distribution (SED) inferred from photometry at the same epoch (see Section~\ref{sec:analysis}), followed by a progressive decrease to $\sim5600\pm600\,\rm{K}$ during the following $\sim30\,\rm{d}$.
Although prominent Balmer lines (in particular \ha~and \hb) and high-ionization features can largely affect the shape of the pseudo-continuum, we do not see a significant improvement in the fit, and we do not get different temperatures excluding regions dominated by narrow emission lines.
On the other hand, the high temperatures inferred during the first $6.1\,\rm{d}$ suggest a SED peaking at bluer wavelengths, not covered by the optical spectra, which in turn can significantly affect the proper determination of the temperature of the pseudo-continuum.
\begin{figure}
\begin{center}
\includegraphics[width=0.975\columnwidth]{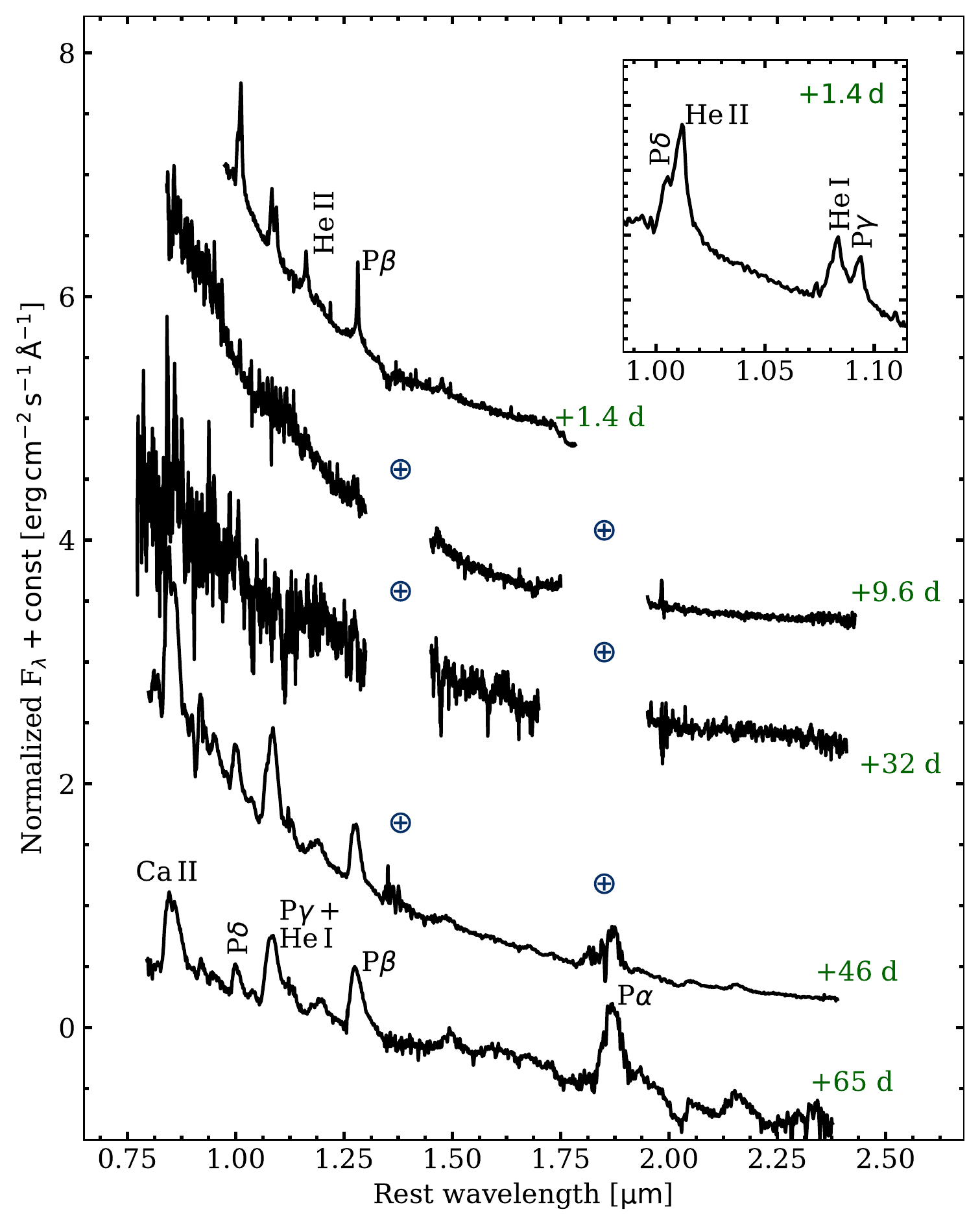}
\caption{NIR spectra of \object. Spectra were corrected for the Galactic and host extinction and wavelengths have been corrected for redshift as inferred from the positions of the narrow \ion{Na}{1D} galactic features. $\oplus$ symbols mark the positions of the most important telluric absorption features. The $+65\,\rm{d}$ spectrum is shown in logarithmic scale in order to facilitate the comparison with earlier spectra. Spectra at $+9.6\,\rm{d}$ and $+32\,\rm{d}$ have been re-binned to $1/3$ of their original resolutions to increase their S/N ratios. 
The inset shows a line identification in the blue part of the spectrum at $+1.4\,\rm{d}$. Spectra used to create this figure are available as Data behind the Figure. \label{fig:NIRspectra}}
\end{center}
\end{figure}

Balmer lines are visible throughout the spectroscopic evolution of \object.
Up to $+4.6\,\rm{d}$, the most prominent H lines (i.e., \ha~and \hb) are purely in emission, with a slightly blue-shifted peak \citep[$V_{shift}\simeq70$\kms, probably due to a `macroscopic' velocity $V_{bulk}$ of the recombining shell; see][]{2014ApJ...797..118F}, characterized by narrow cores with a full--width--at--half--maximum (FWHM) of $\simeq300$\kms, and broader wings extending up to full--width--at--zero--intensity (FWZI) of $\simeq4000$\kms.
Similar velocities are also inferred from the profile of \hb.
While the overall profile might be contaminated by host lines (such as \ion{[N}{2]} $\lambda\lambda6548,6583$), we note that the \ha~is not well reproduced using a single Gaussian or Lorentzian profile.
This might either suggest the presence of recombining shells moving at different velocities, or a broadening due to electron scattering in a dense ionized medium (see Section~\ref{sec:analysis}).
At $+6.1\,\rm{d}$, the flux of the narrow Balmer lines decrease significantly (see above) and broad boxy absorption features appear in the blue part of \ha~and \hb.
Interestingly, we infer different expansion velocities from the absorption minima of \ha~and \hb~ ($V_{\rm{H}\alpha}=6500$\kms~vs $V_{\rm{H}\beta}=1400$\kms), suggesting an intrinsic difference in the expansion velocities of the absorbing shells.
At $t\ge+36\,\rm{d}$, the \ha~region ($6100-6700$\ang) is dominated by a broad, blue-shifted and boxy emission and a broad boxy absorption component, with expansion velocities extending from $\simeq1.2\times10^4$ to $\simeq1.8\times10^4$\kms~with respect to \ha~rest wavelength, becoming progressively fainter and disappearing at $t\gtrsim51\,\rm{d}$.
This can alternatively be identified as \ion{Si}{2} $\lambda6355$, which would result in expansion velocities comparable to those derived from \ion{Fe}{2} $\lambda5169$ (see Section~\ref{sec:analysis}).
This interpretation is also supported by the overall shape of the emission component, symmetric with respect to its centroid and well reproduced using a single Gaussian profile with a FWHM of $\simeq5800$\kms. 
At the same epoch, \ha~also shows a sharp P Cygni profile with a narrow emission component roughly peaking at the line rest-wavelength, with an absorption component extending up to $3\times10^3$\kms, reminiscent of narrow features observed in sufficiently high-resolution spectra of Type IIn SNe \citep[see, e.g.][]{2014ApJ...797..118F,2020A&A...635A..39T}.
A more in-depth analysis also reveals an alternative decomposition, with \ha~being the sum of a narrower component on top of a broader, flat-topped profile, also resulting in the presence of a blue `shoulder', similar to that observed in SN~2013L \citep[see, e.g., Figures 22 and 24 in][]{2020A&A...638A..92T} and other similar interacting transients.
This is also confirmed by the \ha~profile at $+472\,\rm{d}$ (see Figure~\ref{fig:halpha}).

\subsection{Near infrared spectra} \label{sec:NIRspectra}
NIR spectroscopy of \object~was triggered soon after discovery and resulted in a very early observation performed only $1.4\,\rm{d}$ after the estimated explosion epoch.
To our knowledge, this is the earliest NIR spectrum of a Type II SN ever obtained.
The complete NIR spectroscopic follow-up campaign spanned a period of $\sim65\,\rm{d}$ and spectra are shown in Figure~\ref{fig:NIRspectra}, while reduction steps and information about the facilities used are described in Appendix~\ref{sec:specredu}.

At $+1.4\,\rm{d}$, the spectrum shows a blue continuum with prominent narrow Paschen ($\rm{Pa}\beta$ to $\rm{Pa}\delta$) lines, \ion{He}{1} ($\lambda10830$) and \ion{He}{2} ($\lambda10124$ and $\lambda11626$) lines in emission, analogous to the flash features seen in the early optical spectra.
At later times we note a spectroscopic evolution similar to that observed in the optical spectra, with a blue, almost featureless continuum at both $+9.6$ and $+32\,\rm{d}$, with a progressive metamorphosis towards spectroscopic features typical of photospheric phases of Type II SNe.
At $+46\,\rm{d}$, roughly corresponding to the end-point of the plateau phase (see Section~\ref{sec:lightcurves}), we note broader Paschen lines ($\rm{Pa}\alpha$ to $\rm{Pa}\delta$), along with $\rm{Br}\beta$ and \ion{He}{1} $\lambda10830$ features, becoming stronger at $+65\,\rm{d}$ (see Figure~\ref{fig:NIRspectra}).
At this epoch, \ion{He}{1} clearly shows a P Cygni profile, with an absorption component extending up to $13400$\kms, possibly consisting in a high-velocity (HV) component centered at $-8650$\kms, with a redder one centered at $-6300$\kms~with respect to \ion{He}{1} rest wavelength.
We note a similar structure also at $+65\,\rm{d}$, suggesting that this feature is real, although we cannot rule out the contribution of other lines such as P$\gamma$ and \ion{Sr}{2}.
If real, along with the measured EW ($\simeq10$\ang~at $+46\,\rm{d}$), this would suggest a ``weak" classification for \object~when compared to the sample presented in \citet{2019ApJ...887....4D}.
On the other hand, given the lack of the blue ``notch" in the \ion{He}{1} emission component, attributed to P$\gamma$+\ion{Sr}{2} and typically observed in weak SNe, we favor the P$\gamma$+\ion{Sr}{2} identification for the blue feature, which in turn would suggest a ``strong" classification for \object.
This would also confirm the claim that fast declining Type II SNe belong to the strong subclass \citep[see][]{2019ApJ...887....4D}.
\begin{figure}
\begin{center}
\includegraphics[width=\columnwidth]{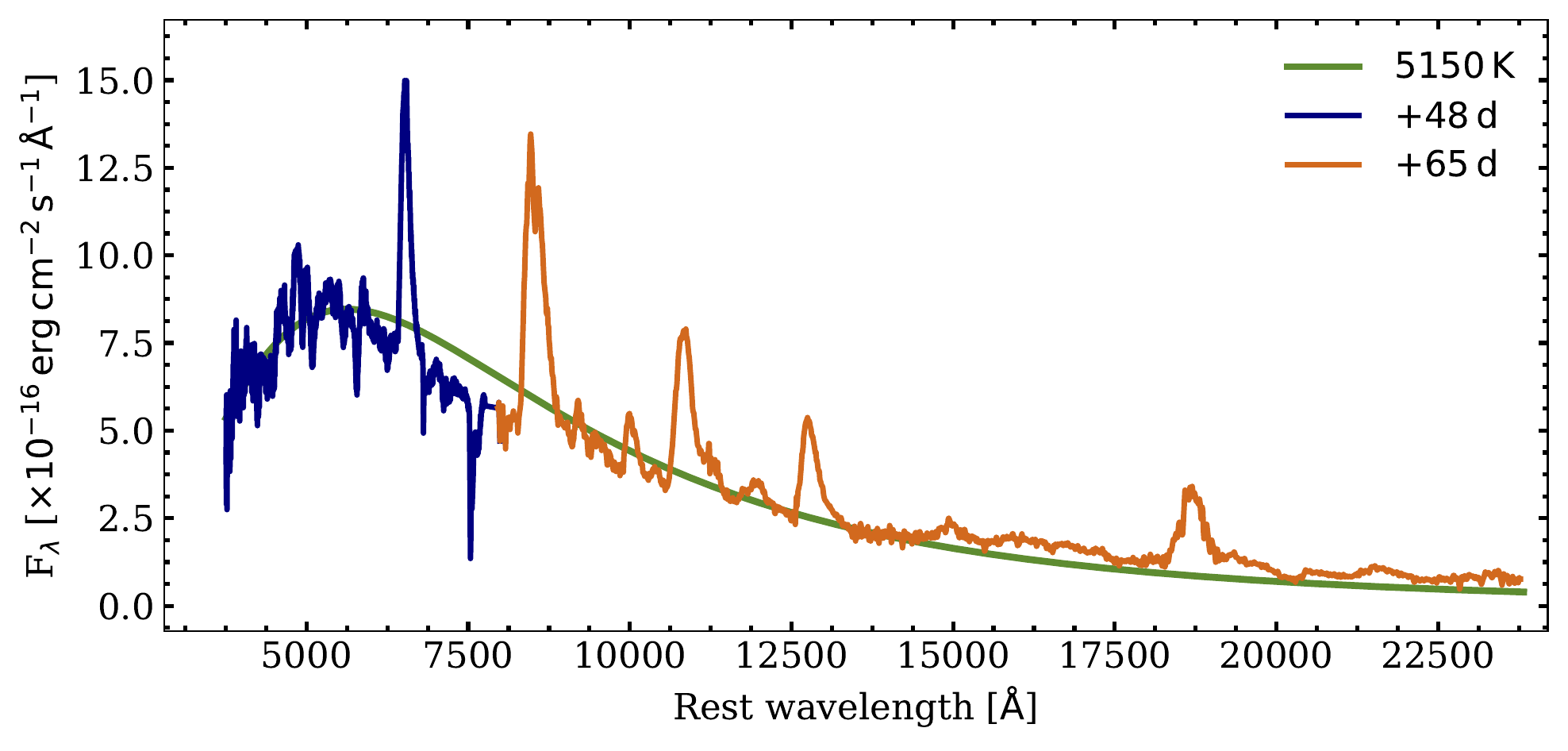}
\caption{Combined optical+NIR spectra of \object~at $+48\,\rm{d}$ and $+65\,\rm{d}$. The spectral continuum is fairly well reproduced by a single BB (although affected by significant line-blanketing at $\lambda\lesssim5000$\ang), ruling out the presence of a colder component at least until $+65\,\rm{d}$ \label{fig:optNIRspec}}
\end{center}
\end{figure}

Isolated narrow lines at $+1.4\,\rm{d}$ (i.e., $\rm{Pa}\beta$ and \ion{He}{2} $\lambda10124$) are fairly well reproduced by a single Lorentzian profile, suggesting that the lack of broader electron scattering wings is most likely due to the lower resolution compared to the $+2.1\,\rm{d}$ and $+4.6\,\rm{d}$ optical spectra (see Figure~\ref{fig:optSpectra} and the description of optical spectra in Section~\ref{sec:optspectra}).
By $t=+9.6\,\rm{d}$, the $\rm{Pa}\beta$ integrated luminosity shows a decrease of $\Delta L=3.3\times10^{38}$\ergs~and also high-ionization \ion{He}{2} lines fade below the level of the spectral continuum.
At $+46\,\rm{d}$ and $+65\,\rm{d}$, Paschen lines show a slightly asymmetric profile, although still reasonably well reproduced by Gaussian profiles with FWHM of $\simeq6500$\kms.

Fitting a BB to the spectral continuum we find an evolution similar to that observed in the optical spectra, although with slightly lower temperatures.
Given the lack of NIR photometric data after peak and the similar spectroscopic evolution to that observed in SN~1998S, we investigate the possible presence of an IR excess by shifting the blue part of the NIR spectra at $t>+9.6\,\rm{d}$ (i.e. the epochs missing a proper absolute flux calibration against photometry) in order to match the red parts of the optical spectra obtained at similar phases.
In addition, we also compute synthetic $z,J,H,K$ magnitudes from the derived spectra, in order to have an estimate of the NIR part of the SED at these epochs (see Section~\ref{sec:analysis}).
In Figure~\ref{fig:optNIRspec} we show the resulting optical+NIR spectrum obtained combining the $+48\,\rm{d}$ optical spectrum with the $+65\,\rm{d}$ NIR one, clearly showing no evidences of a colder component, at least until $+65\,\rm{d}$.

\section{Analysis and discussion} \label{sec:analysis}
In the following, we will derive and discuss the main physical quantities obtained through simple modeling of the main observables described in the previous sections.
In Sections~\ref{sec:CMFGEN} and \ref{sec:specmodels} we will also discuss the results of our numerical modeling of the early evolution of \object, and compare our results to those predicted by hydrodynamical models available in the literature.

\begin{figure}
\begin{center}
\includegraphics[width=\columnwidth]{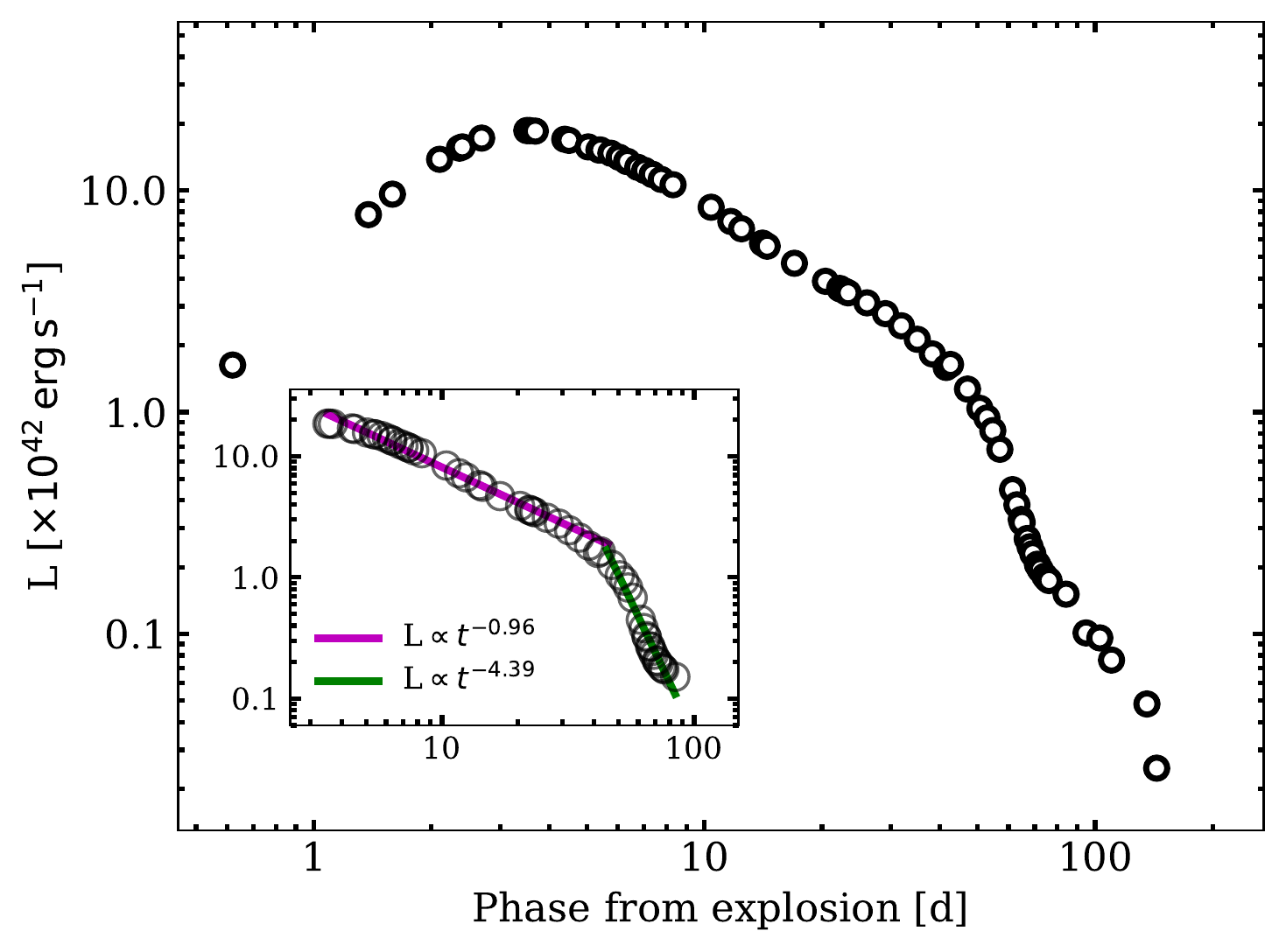}
\caption{Evolution of the pseudo-bolometric luminosity of \object~in logarithmic scale. In the inset, a zoom-in of the region between $+3.5\,\rm{d}$ and $+86\,\rm{d}$, showing the `broken power-law' typical of interacting transients, with an initial evolution described by $L(t)\propto t^{-0.96}$ followed by a steeper decline described by $L(t)\propto t^{-4.39}$ with the break occurring at $t\simeq+45\,\rm{d}$. \label{fig:bolometric}}
\end{center}
\end{figure}
\subsection{Evolution of the bolometric luminosity} \label{sec:bolometric}
The evolution of the bolometric luminosity was estimated following the prescriptions of \citet{2020A&A...635A..39T}, including the contribution to the SED of the early UV bands at $t\lesssim20\,\rm{d}$ and extending the $z,J,H,K$ light curves up to $+65\,\rm{d}$ using synthetic magnitudes obtained from calibrated NIR spectra (see Section~\ref{sec:NIRspectra}).
The derived SEDs at each epoch were integrated using BBs without introducing any `suppression' factor at wavelengths bluer than $\sim3000$\ang~\citep[see, e.g., the discussion in][]{2017ApJ...850...55N}.
This approach is based on a compromise between absorptions due to line blanketing \citep[see, e.g.,][]{2010ApJ...724L..16P,2011ApJ...743..114C} and the UV flux excess predicted by synthetic spectra of Type IIn SNe \citep[see, e.g., Figure~13 and the discussion in][]{2015MNRAS.449.4304D}.
Based on these considerations, the resulting evolution, shown in Figure~\ref{fig:bolometric}, should still be considered as a `pseudo-bolometric' light curve, possibly underestimating the actual luminosity of \object, in particular at early phases.

The bolometric light curve shows a fast rise lasting $\simeq3.7\,\rm{d}$ with a peak luminosity of $\simeq1.9\times10^{43}$\ergs~rapidly declining to $2.5\times10^{40}$\ergs~at $\sim+144\,\rm{d}$.
The corresponding total radiated energy within the first $\sim144\,\rm{d}$ is $\simeq2.3\times10^{49}\,\rm{erg}$.
At $+3.5\,\rm{d}\le t\le+86\,\rm{d}$ the luminosity evolution is well reproduced by a broken power-law, a behavior typically observed in Type IIn SNe and other interacting transients \citep[see, e.g.,][]{2014ApJ...797..118F,2014ApJ...781...42O,2020A&A...635A..39T}. 
We find that the bolometric light curve is well reproduced by $L(t)=7.47\times10^{43}\,t^{-0.96}$\ergs~up to $\simeq+40\,\rm{d}$, followed by a much steeper decline described by $L(t)=3.1\times10^{49}\,t^{-4.39}$\ergs~up to $\sim+86\,\rm{d}$.

The late-time (i.e., during the post-plateau phases) bolometric light curve can be used to infer the mass of the radioactive $^{56}\rm{Ni}$ expelled by the SN explosion.
During the nebular phase, the bolometric light curves of SNe settle onto a ``radioactive" tail, where the energy output is dominated by the ${^{56}\rm{Co}}\rightarrow{^{56}\rm{Fe}}$ decay.
Assuming full $\gamma$-ray trapping (and hence a decline of $\simeq1\,\rm{mag}\,100\,\rm{d^{-1}}$) within the opaque SN ejecta, it is therefore possible to get an estimate of the ejected $^{56}\rm{Ni}$ mass through direct comparison of the late-time luminosity with that of SN~1987A at similar phases, through the relation:
\begin{equation}
M({^{56}\mathrm{Ni}})=0.075 M_{\odot}\times\frac{L_{SN}(t)}{L_{87A}(t)}
\label{eq:}
\end{equation}
\citep[see, e.g.,][and references therein]{2014MNRAS.439.2873S}.
\begin{figure}
\begin{center}
\includegraphics[width=\columnwidth]{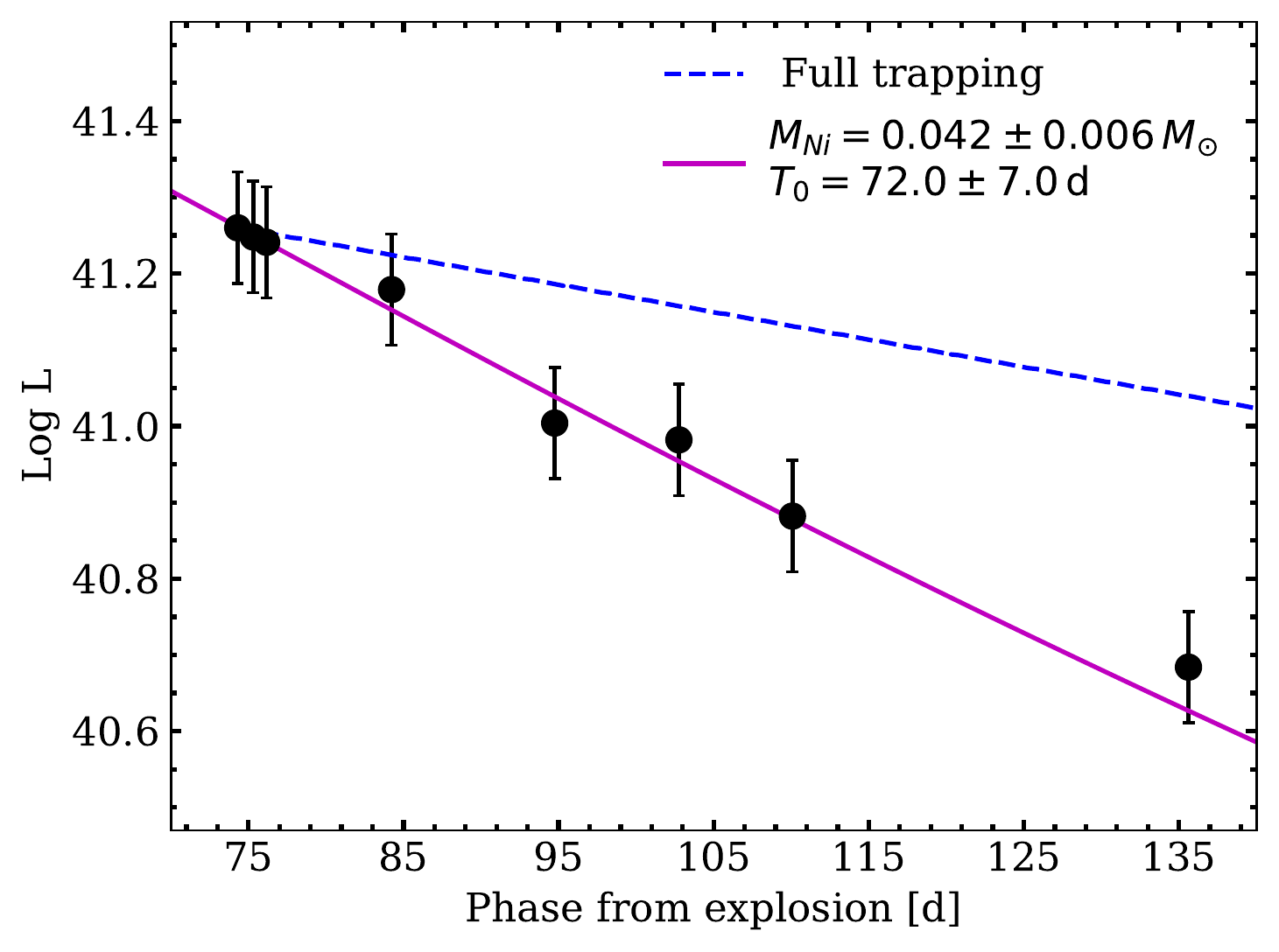}
\caption{Fit of the modified radioactive decay model to the late-time bolometric light curve of \object. The full $\gamma$-ray trapping model is shown as a comparison. \label{fig:nifit}}
\end{center}
\end{figure}
\begin{figure*}
\begin{center}
\includegraphics[width=0.8\linewidth]{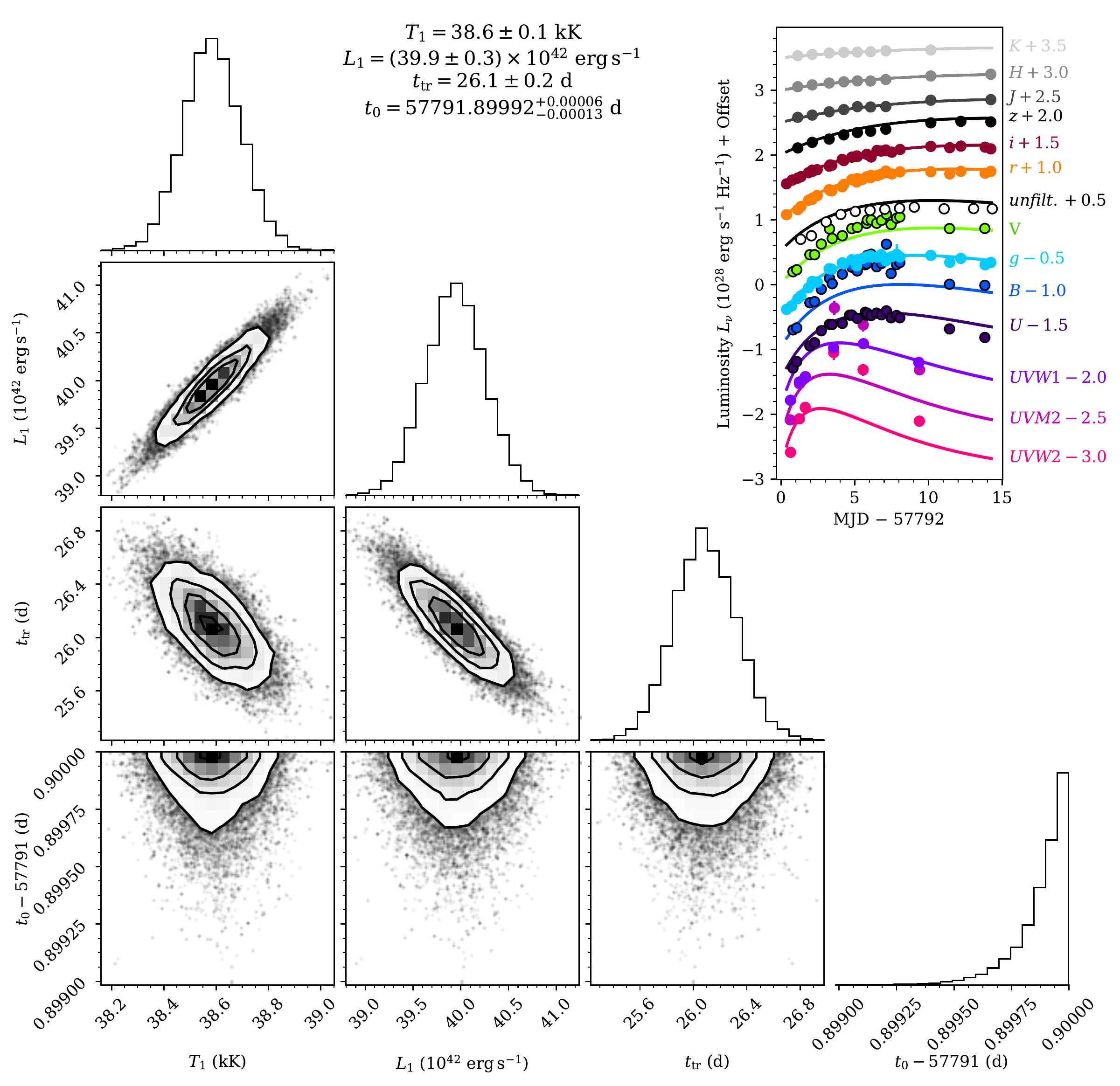}
\caption{Posterior probability distributions and correlations between temperature (T1) and total luminosity (L1) at $+1\,\rm{d}$,
the epoch at which the envelope becomes transparent ($t_{tr}$), and discovery epoch ($\Delta t_0$) following the prescriptions of \citet{2018ApJ...861...63H}. Shock cooling models are not able to reproduce the early light curves of \object.\label{fig:mcmc}}
\end{center}
\end{figure*}
On the other hand, the late-time evolution of \object~(see Figure~\ref{fig:nifit}) shows a much steeper decline if compared to the one expected from the radioactive $^{56}\rm{Co}$ decay.
This is probably due to a non-complete trapping of the $\gamma$-rays produced in the radioactive decay.
A similar evolution was also observed in the Type II-L SN~2014G, also showing high ionization features in early spectra, and attributed to a non-complete trapping of the $\gamma$-rays produced in the $^{56}\rm{Co}$ decay \citep{2016MNRAS.462..137T}.
Non-complete trapping has been discussed by \citet{1997ApJ...491..375C}, who found a simple relation to describe the late-time photometric evolution for a sample of stripped envelope SNe:
\begin{equation}
L(t)=L_0(t)\times\left[1-e^{-(T_0/t)^2}\right]
\label{eq:gammaleak}
\end{equation}
with $T_0$ the full-trapping characteristic time-scale defined as: 
\begin{equation}
T_0=\left(C\kappa_{\gamma}\frac{M^2_{ej}}{E_k}\right)^{1/2},
\label{eq:gammatraptimescale}
\end{equation}
where $M_{ej}$, $E_k$ and $k_{\gamma}$ are the total ejected mass, kinetic energy and the $\gamma$-ray opacity and $C$ a constant given by $C=(\eta-3)^2[8\pi(\eta-1)(\eta-5)]$ for a density profile of the radioactive matter $\rho(r,t)\propto r^{-\eta}(t)$.
The theoretical luminosity due to fully trapped $^{56}\rm{Co}$ energy deposition is given by \citep[see, e.g.,][and references therein]{2012A&A...546A..28J}:
\begin{equation}
L_0(t)=9.92\times10^{41}\frac{M_{^{56}\mathrm{Ni}}}{0.07\,M_{\odot}}\left(e^{-t/111.4}-e^{-7/8.8}\right)\,\mathrm{erg}\,\mathrm{s^{-1}}
\label{eq:jerkstrand}
\end{equation}
where $M_{^{56}\rm{Ni}}$ is the Ni mass expelled by the SN explosion.
This model simply assumes spherical symmetry and homologous expansion of shells with the entire radioactive matter located at the center of the explosion.

Including Equation~\ref{eq:jerkstrand} into Equation \ref{eq:gammaleak}, we then fit the late-time bolometric light curve of \object~to get a rough estimate of the ejected $^{56}\rm{Ni}$ mass and the full-trap characteristic time-scale, performing $10^4$ Monte Carlo simulations, randomly shifting the luminosities within their errors.
The resulting fit, giving $M_{^{56}\rm{Ni}}=0.041\pm0.006$\msun~and $T_0=72\pm7\,\rm{d}$ is shown in Figure~\ref{fig:nifit}.
This is consistent with the median value found by \citet{2019A&A...628A...7A} for a sample of Type II SNe (0.032\msun).

\begin{figure*}
\begin{center}
\includegraphics[width=0.9\textwidth]{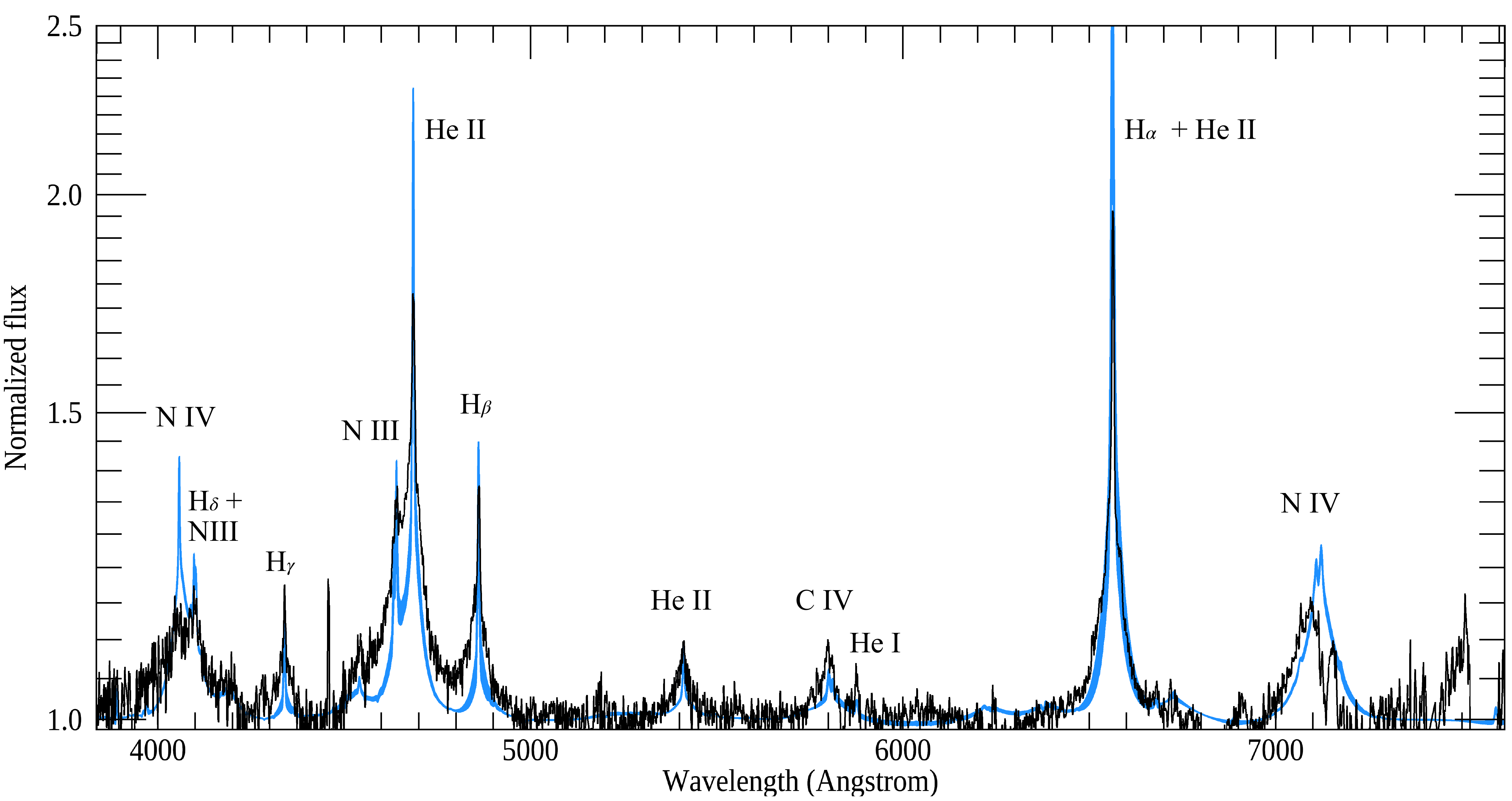}
\caption{Best-fitting CMFGEN models (blue region) compared to the $+2.1$~d optical spectrum of \object.  See Section~\ref{sec:CMFGEN} for our modeling technique and allowed range of parameters. \label{fig:newcmfgenmodel}}
\end{center}
\end{figure*}
\subsection{Shock cooling modeling of the early light curves} \label{sec:radiusFit} 
Theoretical SN models predict a short (seconds to hours) flash of X-ray/UV radiation to be emitted once the radiation mediated shock breaks through the stellar envelope, followed by UV/optical emission from the rapidly expanding and cooling layers.
The analysis of the early post-break cooling phases can be used to infer crucial SN progenitor parameters, including its radius and surface chemical composition \citep[see, e.g.,][]{2017hsn..book..967W}.
In particular, the photospheric temperature and luminosity evolution during the early SN evolution can be described analytically as a function of the shock velocity, the opacity of the expanding medium and the mass and radius of the progenitor star \citep[see, e.g.,][]{2011ApJ...728...63R}.

We model the early photometric evolution of \object~in the context of early SN light curves being dominated by shock cooling radiation escaping from the rapidly expanding progenitor envelope \citep[see][]{2017ApJ...838..130S} using the same approach adopted by \citet{2018ApJ...861...63H}, fitting multi-band light curves assuming $n=3/2$ (polytropic index for a typical RSG envelope) with a Markov Chain Monte Carlo (MCMC) routine with flat priors for all parameters, 100 walkers and 500 steps \citep[see][]{lcfitting}.
The resulting fit (Figure~\ref{fig:mcmc}) shows that the model fails to reproduce the early photometric evolution of \object, which seems to show faster rise times in the UV bands and brighter peaks, in particular in the bluer optical bands.

A plausible explanation must account for an extra source of energy, which, in turn, would affect the accuracy and validity of the \citet{2017ApJ...838..130S} model.
A similar explanation was given by \citet{2018ApJ...861...63H} to explain the fast early evolution of SN~2016bkv.
Interaction with high-velocity SN ejecta and a dense, pre-existing medium, which is typically considered to power the light curves of narrow-lined transients (e.g., SNe IIn; \citealt{1990MNRAS.244..269S} and Ibn; \citealt{2016MNRAS.456..853P}), can affect the overall shape of the light curve (both at early and late phases) and would require a different physical interpretation of the early SN phases.
In stellar explosions occurring within a dense CSM, in fact, the SN shock is expected to break through the dense CSM surrounding the progenitor star rather than the stellar envelope, extending and diluting the SN radiation, with early light curves being dominated by photon diffusion rather than shock-cooling emission.

Although narrow lines are generally considered an indirect proof of ongoing interaction between expanding SN ejecta and a dense pre-existing CSM, high-ionization features (\ion{C}{4}, \ion{N}{3} and \ion{N}{4}) are typically observed only at the very early phases (hours to a few days after explosion) and are believed to arise from the recombining CSM ionized by the shock breakout flash, rather than by photons emitted in shocked regions.
On the other hand, in \object, such features disappear $\simeq6\,\rm{d}$ after explosion, suggesting a simple scenario where the recombining CSM is progressively swept up by the rapidly expanding ejecta. 
Under specific conditions, an efficient conversion of kinetic energy into radiation would therefore be able to provide the required energy input to explain the early evolution of \object.
This is also in agreement with the results obtained by \citet[][see also \citealt{2018ApJ...858...15M}]{2017ApJ...838...28M} modeling the light curves of fast-declining Type II SNe, suggesting red supergiants surrounded by a dense CSM as viable progenitors and that the presence of such dense medium might be common among H-rich CC SNe.
Although the pseudo bolometric light curve of \object~does not show the sudden drop at $t\simeq+25\,\rm{d}$ predicted by the models of \citet[][corresponding to the dense shell becoming optically thin and the photosphere receding into the SN ejecta, see, e.g., their Figure~4]{2011MNRAS.415..199M}, its overall shape is similar to their $10^{-3}$\msunyr~model (see Sections~\ref{sec:CMFGEN} and \ref{sec:specmodels}).

Interaction would also explain narrow \ha~components with sharp P Cygni profiles at later times, observed up to $+36\,\rm{d}$ and the prominent boxy profile observed in the \ha~profile at $t>+36\,\rm{d}$ (Figure~\ref{fig:halpha}), showing a progressively asymmetric profile with a characteristic blue-shifted ($V_{shift}\simeq5000$\kms) ``shoulder".
Similar features are common among Type II SNe showing linearly declining light curves (see, e.g., the cases of SNe~1999ga \citealt{2009A&A...500.1013P} and 2017ivv \citealt{2020arXiv200809628G} and the objects in the sample of \citealt{2014MNRAS.445..554F}) as well as in a few Type IIn SNe at sufficiently late times (see, e.g., the cases of SNe~2005ip \citealt{2012ApJ...756..173S,2017MNRAS.466.3021S} and 2013L \citealt{2020A&A...638A..92T}) and are typically considered evidence of emission from a shocked thick shell of gas \citep[see, e.g.,][]{2017hsn..book..795J}.
Ongoing interaction of SN ejecta with a dense CSM at $t\gtrsim+6\,\rm{d}$ (i.e., when high-ionization features seem to disappear) is also consistent with the radio non-detection of \object~at 5.5 and $9\,\rm{GHz}$ \citep{2017ATel10147....1R}, suggesting efficient synchrotron self absorptions by free electrons in a dense medium at $+21\,\rm{d}$.
\begin{figure*}
\begin{center}
\includegraphics[width=0.9\linewidth]{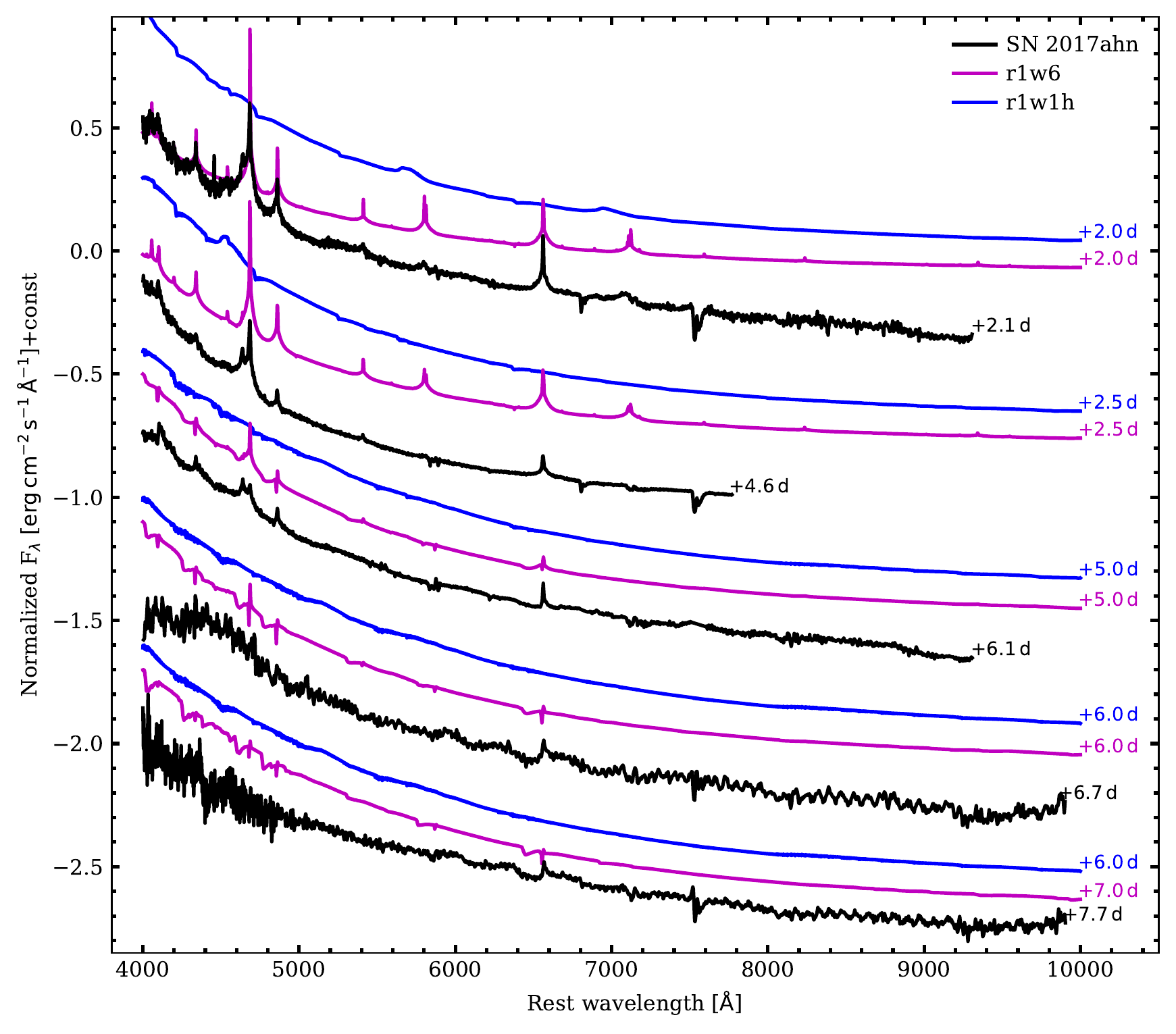}
\caption{Comparison of the early ($t\lesssim8\,\rm{d}$) spectra of \object, corrected for the total reddening reported in Section~\ref{sec:host}, with a selection of \citet{2017A&A...605A..83D} models.
Model r1w1h reproduces well the shape of the continuum at all phases, although it does not show emission features observed in the early spectra of \object. Model r1w6 is able to reproduce both the spectral continuum and most of the emission lines at $t>+2.5\,\rm{d}$, while at $+2\,\rm{d}$ the overall spectral features are not reproduced well by none of the models presented in \citet{2017A&A...605A..83D}. 
The choice of these models is based on their similar evolutions with respect to \object, including the presence of high-ionization features (e.g., \ion{Ne}{3}, \ion{He}{2}, \ion{C}{3,IV}) at early phases, the shape of the continuum and the timescale at which the spectra evolve (e.g., all other models have a much faster evolution, with high-ionization features disappearing a few hours after the shock-breakout; see Figures~C.1-C.7 in \citealt{2017A&A...605A..83D}).
These both correspond to a progenitor radius $R_{\star}=501$\rsun, with mass-loss rates $10^{-6}$ (r1w1h) and $10^{-2}\,\rm{M_{\odot}}\,\rm{yr^{-1}}$ (r1w6, with $\dot{M}=10^{-6}\,\rm{M_{\odot}}\,\rm{yr^{-1}}$ beyond $5\times10^{14}\,\rm{cm}$, see the main text and \citet{2017A&A...605A..83D} for more details). \label{fig:DessartSpectra}}
\end{center}
\end{figure*}

\subsection{CMFGEN models of the early interaction} \label{sec:CMFGEN}
To investigate the properties of \object~at early times, we compute numerical models using the radiative transfer code CMFGEN \citep{hm98} with the implementation of \citet{groh2014a}. 
In our models, photons diffuse out through the extended material around the progenitor. 
Although we do not need to assume a source of energy, the photons are thought to be produced by the interaction between the SN ejecta and the progenitor wind. 
The wind heats up and emits continuum and line photons according to its temperature and density structure. 
Our main assumptions are of a spherical and stationary wind, radiative equilibrium to compute the temperature structure, non-local thermodynamic equilibrium, and time-independent radiative transfer. 
The models take as input the luminosity of the event $L_{\rm SN}$, the progenitor mass-loss rate \mdot, wind terminal velocity $\upsilon_{\infty}$ and surface abundances, and the location of the inner boundary $R_{in}$, which depends on the ejecta dynamics and time after explosion. 
Since the progenitor wind is dense, the photosphere is extended and we quote two flux temperatures, $T_{\star}$ at the inner boundary (Rosseland optical depth of $\tau\simeq10.0$), and $T_{\rm eff}$ at $\tau=2/3$.
We refer the reader to \citet{BoianGroh2018}, \citet{BoianGroh2019} and \citet{BoianGroh2020} for further details.

We use the models presented  in \citet{BoianGroh2020} as a starting point. 
These model grids were computed at 1.0, 1.8, and $3.8\,\rm{d}$ after explosion, exploring  a wide range of parameters: $1.9\times10^8\le L\le2.5\times10^{10}$\lsun, $5\times10^{-4}\le\dot{M}\le10^{-2}\,\rm{M_{\odot}}\,\rm{yr^{-1}}$, three values of chemical abundances (solar, He-rich and CNO-processed) and fixed wind terminal velocities and ejecta expansion velocities of $\upsilon_{\infty}=1.5\times10^2$\kms~and $v_{ej}=10^4$\kms, respectively.
Once we find the best fit models among the \citet{BoianGroh2020} grid, we compute 30 additional models in this region of parameter space to obtain the properties of \object~at $+2.1\,\rm{d}$, corresponding to our highest S/N optical spectrum around that epoch.
Figure~\ref{fig:newcmfgenmodel} shows our best fitting models compared with the observed spectrum.

We find that our CMFGEN models quantitatively reproduce the spectral morphology of \object~at $+2.1\,\rm{d}$, with strong \ion{He}{2}, \ion{H}{1}, \ion{N}{3}, and \ion{N}{4} features.
Our models indicate $L=5.0-7.2\times10^9$\lsun, $\dot{M}=2.7-4.0\times10^{-3}$\msunyr~(for $\upsilon_{\infty}=150$\kms), $R_{in}=2.17\times10^{13}\,\rm{cm}$, $T_{\star}=26600-28900\,\rm{K}$, surface mass fractions $C_{\rm sur}=5.6\times10^{-5}$, $N_{\rm sur}=8.2\times10^{-3}$, and $O_{\rm sur}=1.3\times10^{-4}$, with an estimated 3-$\sigma$ error of 50\% and a He surface mass fraction of $Y\simeq0.35-0.50$, suggesting some He enhancement, consistent with the scenario that the progenitor lost a significant fraction of its H envelope before exploding.
Fitting the observed SED implies a total color excess of $E(B-V)=0.06\pm 0.01\,\rm{mag}$ and $R_V=3.1$, which, although in line with the extinction values found for a large sample of interacting SN \citep{BoianGroh2020}, is in contrast with the reddening derived from both the optical spectra (clearly showing strong \ion{Na}{1D} features at all times; see Section~\ref{sec:optspectra}) and the spectral analysis of the local environment (Section~\ref{sec:host}).

Our results show that the progenitor had CNO-processed surface abundances, with enhanced N and depleted C and O. 
The Geneva stellar evolution models suggest that this would be expected from a massive ($\sim15-25$\msun) RSG, a YHG or a BSG/LBV star \citep{groh2013c}. 
However, these models make strong assumptions about rotational mixing and mass loss, which have a key impact on the final CNO surface abundances \citep{meynet2013,meynet2015}. 
In addition, a significant fraction of massive stars evolve in binary systems \citep{sana2012}, with important consequences for the final mass and surface abundances. 
Finally, the final mass of red supergiants could be much larger than previously thought \citep{Farrell2020a,Farrell2020b}. 
For all these reasons, it is challenging to map final surface abundances to progenitor initial or final masses \citep[see also the discussion in][]{BoianGroh2020}.

We remark that the above quantities should be taken with caution given our model assumptions. 
In particular, an important role may also be played by asymmetries and/or particular geometrical configurations of the CSM, which were also believed to affect the evolution of the observables of SN~1998S \citep[see][]{2000ApJ...536..239L} as well as CC SNe in general. 
We encourage further constraints on the CSM morphology of interacting SNe, which would allow these effects to be taken into account in future modeling of the early evolution of \object.
\begin{figure}
\begin{center}
\includegraphics[width=\columnwidth]{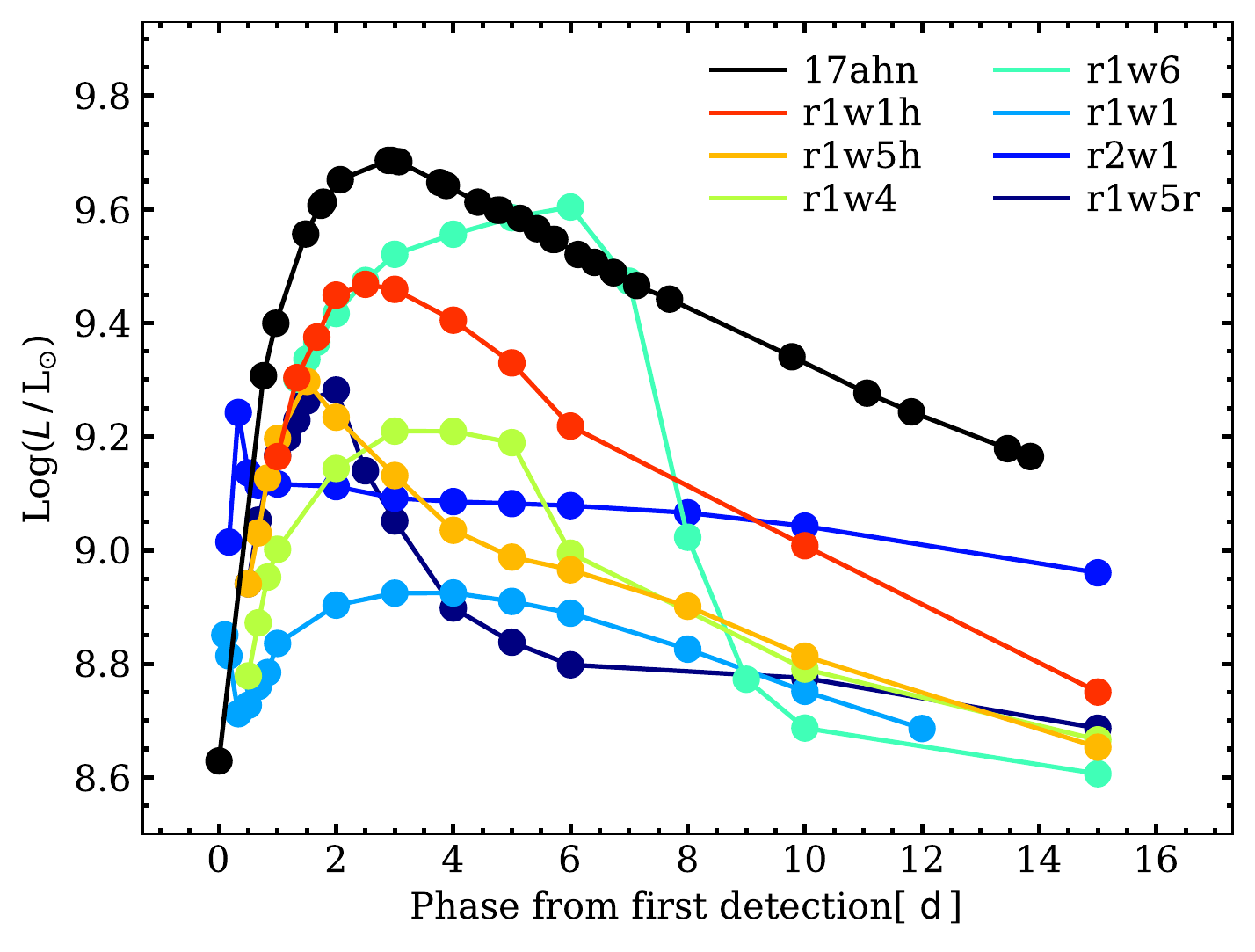}
\caption{Comparison of the pseudo-bolometric light curve of \object~with the {\it uvoir} models of \citet{2017A&A...605A..83D}. Model r1w1h, corresponding to a progenitor radius $R_{\star}=501$\rsun~with a mass-loss rate $\dot{M}=10^{-6}\,\rm{M_{\odot}}\,\rm{yr^{-1}}$ and an atmospheric density scale height $H_{\rho}=0.1R_{\star}$ ($H_{\rho}=0.3R_{\star}$ down to $\rho=10^{-12}\,\rm{g}\,\rm{cm^{-3}}$; see the main text and \citealt{2017A&A...605A..83D} for details) reproduces well the shape of the light curve, although with a slightly fainter luminosity at all phases. \label{fig:DessartUVOIR}}
\end{center}
\end{figure}

\subsection{Comparison with existing hydrodynamic models} \label{sec:specmodels}
After investigating our own models, we now compare the early photometric and spectroscopic evolution of \object~with existing hydrodynamical models available in the literature.
\citet{2017A&A...605A..83D} used 1-D radiation-hydrodynamics and 1-D non-local thermodynamical equilibrium (LTE) radiative transfer models to reproduce photometric and spectroscopic features of RSG stars exploding within moderately extended and massive winds, with $R_w\sim10\,R_{\star}$ and a total mass $\lesssim10$\msun.
The resulting models were obtained using multi-group radiation-hydrodynamics simulations performed with {\sc heracles}\footnote{\url{http://irfu.cea.fr/Projets/Site_heracles/}} \citep{2007A&A...464..429G} and post-processed using the radiative transfer code {\sc cmfgen}\footnote{\url{http://kookaburra.phyast.pitt.edu/hillier/web/CMFGEN.htm}} \citep{1998ApJ...496..407H} with initial conditions described in \citet{2013MNRAS.433.1745D,2015MNRAS.449.4304D}.
The result of a simple direct comparison of the observed early evolution of \object~with their synthetic spectra is shown in Figure~\ref{fig:DessartSpectra}.

Although the evolution of the narrow spectral features is well matched by their r1w6 model, it fails to reproduce the shape of the blue spectral continuum, while the temperature of the pseudo photosphere is well matched by model r1w1h, which, on the other hand, is not able to reproduce the evolution of the high ionization features.
These models both correspond to a RSG progenitor with $R_{\star}=501$\rsun, a total ejected mass of 12.52\msun~with kinetic energy $E_k=1.35\times10^{51}\,\rm{erg}$, colliding with a pre-existing confined wind extending from $10^{15}$ to $2\times10^{16}\,\rm{cm}$, a total mass of 2.89\msun~\citep[based on the mass of the CSM inferred by][for SN~2010jl]{2014ApJ...797..118F} and an expansion velocity $u_w=10^2$\kms.
The model r1w6, in particular, corresponds to a mass-loss rate of $10^{-2}\,\rm{M_{\odot}}\,\rm{yr^{-1}}$ ($10^{-6}$ beyond $5\times10^{14}\,\rm{cm}$), while r1w1h assumes $\dot{M}=10^{-6}\,\rm{M_{\odot}}\,\rm{yr^{-1}}$ with a density scale height $H_{\rho}=0.3R_{\star}$ followed by a power-law with index 12 above $10^{-12}\,\rm{g}\,\rm{cm^{-3}}$ \citep[$H_{\rho}=0.1R_{\star}$; see][]{2017A&A...605A..83D}.
Model r1w1h also seems to reproduce well the shape of the pseudo-bolometric light curve of \object~(see Figure~\ref{fig:DessartUVOIR}, where we compare it with the $uvoir$ light curves\footnote{available at \url{https://www-n.oca.eu/supernova/early/early.html}} from \citealt{2017A&A...605A..83D}), although with lower luminosities at all phases.

The main difference of r1w1 and r1w1h is the adopted value for the mass-loss rate ($10^{-2}$ and $10^{-6}\,\rm{M_{\odot}}\,\rm{yr^{-1}}$ for the dense and weak-wind models r1w6 and r1w1h, respectively), suggesting an intermediate value for \object.
This is in agreement with our modeling discussed in Section~\ref{sec:CMFGEN}, which gives a mass-loss rate of $2.7-4.0\times10^{-3}$\msunyr.
This idea is also supported by the mass-loss rate inferred by \citet{2015ApJ...806..213S} from their modeling of the spectra of SN~1998S, resulting in $\dot{M}=6\times10^{-3}\,\rm{M_{\odot}}\,\rm{yr^{-1}}$.
On the other hand, we cannot rule out other explanations, as well as a combination of different parameters, including masses, velocities and kinetic energies of the expanding SN ejecta and the pre-existing CSM, or a different rate of conversion of kinetic energy into radiation.
\citet{2015MNRAS.449.4304D}, for example, showed that a higher kinetic energy of the SN ejecta can give higher luminosities without affecting the overall shape of the light curve.
Despite the limitations of our approach, the spectral and luminosity evolution predicted by models r1w6 and r1w1h are fairly in agreement with the observed evolution of \object, considering these models were not specifically constructed on its observables.

\begin{figure}
\begin{center}
\includegraphics[width=\columnwidth]{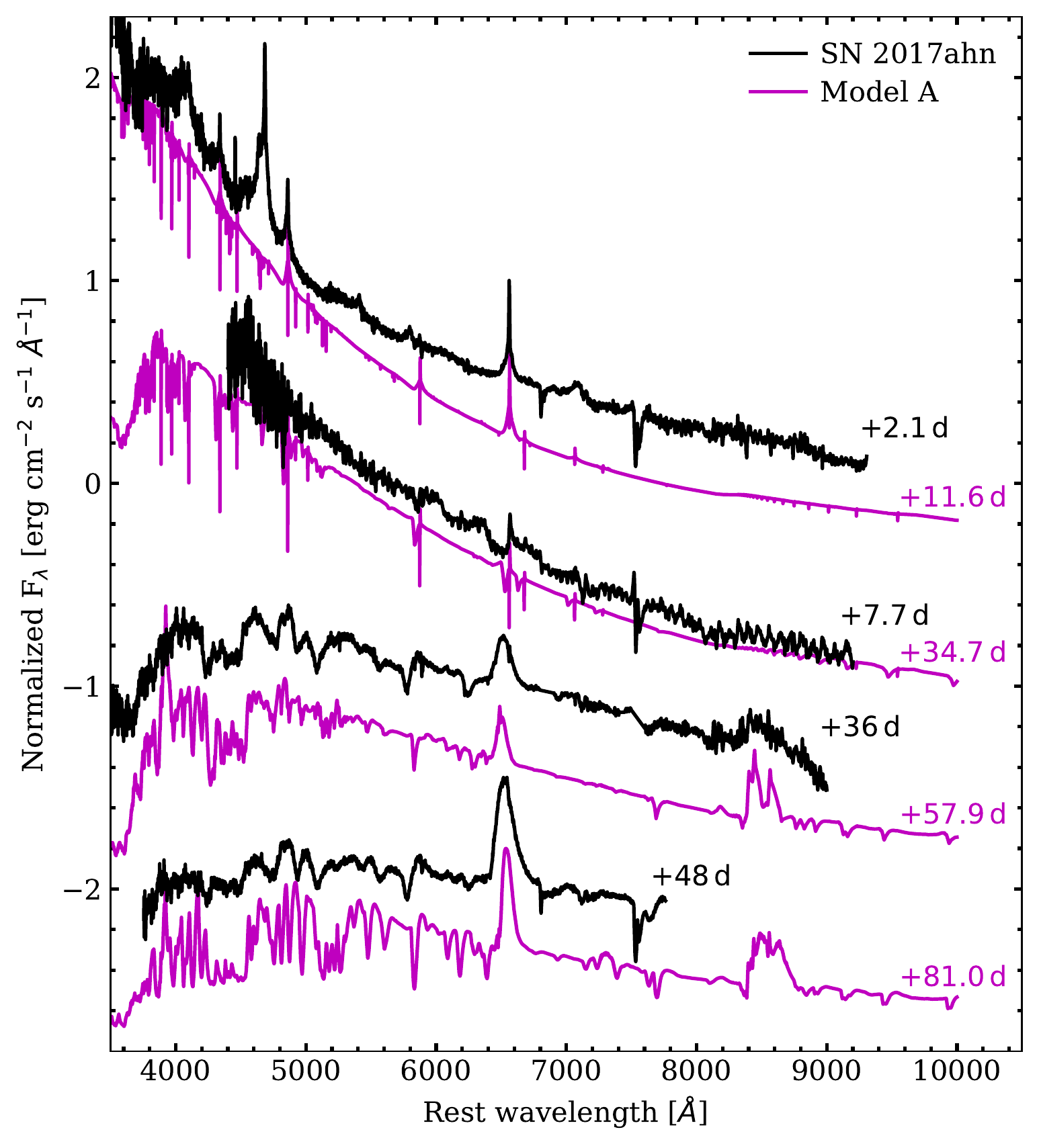}
\caption{Comparison of the spectroscopic evolution of \object~at selected epochs with ``Model A" computed for SN~1998S in \citet{2016MNRAS.458.2094D}. \label{fig:dessart98S}}
\end{center}
\end{figure}

Based on their similar spectroscopic evolutions, we also compare the spectroscopic evolution of \object~with the radiative transfer simulated spectra for SN~1998S presented in \citet{2016MNRAS.458.2094D}.
Although the evolution of the narrow features seems to be slower~(see Figure~\ref{fig:dessart98S}), Model A \citep[see Figure~5 in][]{2016MNRAS.458.2094D} is able to reproduce well both the shape of the spectral continuum and the spectral features observed in \object, although it fails to reproduce the \ion{N}{3}+\ion{He}{2} feature observed at $\le+6.1\,\rm{d}$.
This model, similar to that proposed by \citet{2004MNRAS.352.1213C} for the Type IIn SN~1994W, corresponds to a massive ($M_{ej}=10$\msun) and energetic $E_k=10^{51}\,\rm{erg}$ shell ramming into a 0.4\msun~dense CSM, produced by a stationary wind with a mass-loss rate of 0.1\msunyr~with an expansion velocity of $10^2$\kms~\citep[see also Appendix A of][for more details about the model]{2016MNRAS.458.2094D}.
These parameters are in contradiction with those inferred from simple comparison with other pre-existing models (see above), highlighting the need of a more accurate modeling.

\section{Summary and conclusions} \label{sec:conclusions}
In this paper, we discussed the photometric and spectroscopic properties of the Type II \object, discovered soon after explosion by the DLT40 survey in the nearby galaxy NGC~3318.
Multi-wavelength follow-up campaigns, promptly activated after discovery, revealed a relatively fast early photometric evolution, showing peculiar fast rising light curves in UV bands, suggesting a very high initial temperature of the pseudo photosphere.

Like in SN~1998S, the early spectral evolution is characterized by the presence of prominent and narrow high-ionization features with narrow unresolved cores and broad wings.
These are typical of electron scattering profiles observed in Type IIn SNe, and are expected to be common during the very early phases of the evolution of CC SNe.
These features get progressively fainter and disappear roughly a week after explosion, suggesting the presence of a confined dense shell, progressively swept up by the expanding SN ejecta.

At later times, the evolution of \object~resembles those typically observed in fast declining Type II SNe, with a short {plateau-like} phase lasting $\sim50\,\rm{d}$, followed by a tail similar, although slightly steeper, to that predicted by the $^{56}\rm{Co}$ radioactive decay during the nebular phases of SNe.
According to \citet{2014MNRAS.445..554F}, the rapid decline rates observed in objects like \object~both during the {\it plateau} and the nebular phases are consistent with those expected by fast declining Type II SNe. 
Comparing the late evolution of the pseudo bolometric light curve of \object~to that of SN~1987A, we estimate an ejected $^{56}\,\rm{Ni}$ mass of $0.041\pm0.006$\msun, with a non-complete trapping of the $\gamma-$rays produced in the radioactive Co decay, although this measurement could be affected by ongoing ejecta-CSM interaction.

A non-negligible contribution of the interaction is also suggested by the evolution of the bolometric luminosity showing a {\it broken power-law} shape just before settling on the radioactive tail, while \ha~shows a late-time structured profile with a peculiar blue shoulder consistent with a boxy flat-topped emission component. 
In addition, the poor fit of the early light curves to shock-cooling models might also reveal a significant contribution of interaction to the total luminosity already at early times.

Numerical modeling of the early evolution of \object~using the radiative transfer {\sc cmfgen} \citep{1998ApJ...496..407H} code suggests a massive ($15-25$\msun) progenitor for \object, with a initial radius of $\simeq310$\rsun~and a mass-loss rate $\dot{M}=2.7-4.0\times10^{-3}\,\rm{M_{\odot}}\,\rm{yr^{-1}}$ (assuming $\upsilon_{\infty}=1.5\times10^2$\kms).
A similar result was obtained through direct comparison with existing non-LTE radiative-transfer models obtained with the {\sc heracles} \citep{2007A&A...464..429G} and {\sc cmfgen} codes, suggesting a progenitor with a plausible mass-loss rate of $\simeq3\times10^{-3}\,\rm{M_{\odot}}\,\rm{yr^{-1}}$, although these models assume a less massive RSG progenitors (with an intial mass of $\simeq14$\msun) and terminal velocity \citep[$\upsilon_{\infty}=50$\kms, see][]{2013MNRAS.433.1745D,2017A&A...605A..83D}.
Although both approaches can give models matching many features observed in \object, none of them is able to account for complex geometrical configurations, binarity of the progenitor system and their consequences on the final masses and surface abundances, or reproduce completely the early evolution of the observables, suggesting the need for continued modeling efforts.
Nonetheless, the data and models support the idea that linearly declining Type II SNe arise from massive stars depleted of a significant fraction of their H-rich envelope in the pre-SN stage.

\acknowledgments
We gratefully acknowledge useful comments from N. Morrell.

Based on observations collected at:
The Gemini Observatory, under program GN-2016B-Q-57, which is operated by the Association of Universities for Research in Astronomy, Inc., under a cooperative agreement with the NSF on behalf of the Gemini partnership: the National Science Foundation (United States), the National Research Council (Canada), CONICYT (Chile), Ministerio de Ciencia, Tecnolog\'{i}a e Innovaci\'{o}n Productiva (Argentina), and Minist\'{e}rio da Ci\^{e}ncia, Tecnologia e Inova\c{c}\~{a}o (Brazil). \\
VLT with MUSE, under the ESO programmes 095.D-0172 and 0101.D-0748. \\
The Magellan I Baade telescope at Las Campanas Observatory. \\
The SALT observations presented here were made through Rutgers University programs 2016-1-MLT-007 (PI: Jha); supernova research at Rutgers is supported by NSF award AST-1615455. \\
Part of the funding for GROND (both hardware as well as personnel) was generously granted from the Leibniz-Prize to Prof. G. Hasinger (DFG grant HA 1850/28-1). \\
This work makes use of observations from the Las Cumbres Observatory network of telescopes. \\
SNOoPy is a package for SN photometry using PSF fitting and/or template subtraction developed by E.~Cappellaro. A package description can be found at \url{http://sngroup.oapd.inaf.it/snoopy.html}. \\
L.T. acknowledges support from MIUR (PRIN 2017 grant 20179ZF5KS). \\
Research by D.J.S. is supported by NSF grants AST-1821967, 1821987, 1813708, 1813466, 1908972, and by the Heising-Simons Foundation under grant \#2020-1864.\\
J.S. acknowledges support from the Packard Foundation.
R.C. and M.S. acknowledge support from STFC grant ST/L000679/1 and EU/FP7-ERC grant no [615929].
T.-W.C. acknowledges the EU Funding under Marie Sk\l{}odowska-Curie grant agreement No 842471.
H.K. was funded by the Academy of Finland projects 324504 and 328898. 
D.A.H., J.B., and D.H. are supported by NSF grant AST-1911225 and NASA Swift grant 80NSSC19k1639. 
L.G. was funded by the European Union's Horizon 2020 research and innovation programme under the Marie Sk\l{}odowska-Curie grant agreement No. 839090. This work has been partially supported by the Spanish grant PGC2018-095317-B-C21 within the European Funds for Regional Development (FEDER). \\
D.A.H., C.M., and G.H. are supported by NSF grant 1313484. \\
E.Y.H. acknowledges the support provided by the National Science Foundation under Grant No. AST-1008343, AST-1613472 and AST-1613426.
 
This research has made use of the NASA/IPAC Extragalactic Database (NED), which is operated by the Jet Propulsion Laboratory, California Institute of Technology, under contract with the National Aeronautics and Space Administration. \\
SNOoPy is a package for SN photometry using PSF fitting and/or template subtraction developed by E. Cappellaro. A package description can be found at \url{http://sngroup.oapd.inaf.it/ecsnoopy.html.} \\

\facilities{Las Cumbres Observatory network of telescopes (Sinistro), FTN (FLOYDS), Swift (UVOT), VLT:Yepun (MUSE), SALT (RSS), IRTF (Spex), SOAR (Goodman), Gemini (FLAMINGOS-2), Magellan:BAADE (FIRE), MPG $2.2\,\rm{m}$ telescope (GROND)}

\software{\texttt{astropy} \citep{2013A&A...558A..33A,astropy}, \texttt{SNOoPY}: \url{http://graspa.oapd.inaf.it/snoopy.html}, \texttt{lpipe} \citep{2019PASP..131h4503P}, \texttt{PySALT} \citep{2010SPIE.7737E..25C}, {\sc firehose} \citep{2013PASP..125..270S}, {\sc lightcurve\_fitting} \citep{lcfitting}}

\appendix

\section{Photometric data} \label{sec:photoredu}
PROMPT5 unfiltered DLT40 images were reduced as in \citet{2018ApJ...853...62T}, using our dedicated pipeline and calibrated to the $r$-band.
Magnitudes are reported in Table~\ref{table:prompt}.  

Optical photometry of \object~was obtained mostly using the facilities of the Las Cumbres Observatory network \citep{2013PASP..125.1031B} within the Supernova Key Project.
Additional optical and NIR photometry was obtained, as part of the GREAT survey \citep{2018ApJ...867L..31C}, using the Gamma-Ray Burst Optical/Near-Infrared Detector \citep[GROND;][]{2008PASP..120..405G}, a 7-channel imager, collecting multi-color photometry in $g',r',i',z'$ and $J,H,K_{\rm{S}}$ bands simultaneously, mounted at the $2.2\,\rm{m}$ MPG telescope at ESO La Silla Observatory in Chile. 
These frames were pre-reduced using the dedicated GROND pipeline \citep{2008ApJ...685..376K}, including bias and flat-field corrections, image stacking and astrometric calibrations.
Both optical and NIR magnitudes were obtained using the dedicated pipeline {\sc SNOoPY}\footnote{\url{http://graspa.oapd.inaf.it/snoopy.html}} and are listed in Tables~\ref{table:UBVphot}, \ref{table:grizphot} and \ref{table:JHKphot}.

{\it Swift} data were reduced using the pipeline of the Swift Optical Ultraviolet Supernova Archive \citep[SOUSA;][]{2014Ap&SS.354...89B}, which is based on the method of \citet{2009AJ....137.4517B} using the zero points of \citet{2011AIPC.1358..373B}.
The uncertainties account for differences in the measured brightness using a range of background regions with galaxy light similar to that of the SN. 
Magnitudes are reported in Table~\ref{table:UVOTphot}

\section{Spectroscopic data} \label{sec:specredu}
Optical spectra of \object~were obtained using the facilities listed in Table~\ref{table:speclog}.
Most of the spectra were obtained using the Faulkes Telescopes $2\,\rm{m}$ telescopes of the Las Cumbres Observatory network, located at the Haleakala Observatories (Faulkes Telescope North, Hawaii -- U.S.; FTN) and the Siding Spring Observatory (Faulkes Telescope South, Australia; FTS) using the cross-dispersed, low resolution spectrographs FLOYDS and reduced using their dedicated pipeline\footnote{\url{https://lco.global/documentation/data/floyds-pipeline/}} \citep{Valenti14}.
Optical spectra were also obtained using the $10\,\rm{m}$ Southern African Large Telescope (SALT), located at the South African Astronomical Observatory (SAAO), Sutherland, South Africa, with the Robert Stobie Spectrograph (RSS) \citep[reduced using the dedicate pipeline {\sc PySALT;}][]{2010SPIE.7737E..25C} and the $4.1\,\rm{m}$ SOuthern Astrophysical Research telescope (SOAR) with the Goodman High Throughput Spectrograph \citep{2004SPIE.5492..331C}, located at the Cerro Tololo Inter-American Observatory, Cerro Pach\'on, Chile and reduced using a dedicated pipeline\footnote{\url{http://www.ctio.noao.edu/soar/content/goodman-data-reduction-pipeline}}.
One additional optical spectrum was obtained using the $10\,\rm{m}$ Keck I telescope located at the Mauna Kea Observatories (Hawaii, U.S.) using the Low Resolution Imaging Spectrometer \citep[LRIS;][]{1995PASP..107..375O,2010SPIE.7735E..0RR} and reduced using the automated pipeline {\sc lpipe\footnote{\url{https://www.astro.caltech.edu/~dperley/programs/lpipe.html}}} \citep{2019PASP..131h4503P}.
MUSE data reduction was performed using the ESO MUSE data reduction pipeline under the Reflex interface \citep{2013A&A...559A..96F}, which includes bias subtraction, flat-fielding, wavelength and flux calibrations, background sky subtraction and atmospheric effects correction. 
The spectra were subsequently extracted from the reduced MUSE datacube for further analysis. 

Near-infrared (NIR) spectra were taken with the FLAMINGOS-2 instrument \citep[F2;][]{Eikenberry2006} at Gemini South Observatory, SpeX instrument \citep{Rayner03} on the NASA Infrared Telescope Facility and the Folded-port InfraRed Echellette instrument \citep[FIRE;][]{Simcoe2013} on the Magellan Baade telescope. The IRTF SpeX data was taken in cross-dispersed SXD mode with the 0.5 arcsec slit, yielding wavelength coverage from $\sim0.7-2.4\,\rm{\mu m}$ and a $R\sim1200$.
The Magellan FIRE spectra were obtained in high throughput prism mode with a 0.6\arcsec~slit, giving continuous wavelength coverage from 0.8 to $2.5\,\rm{\mu m}$. 
For the Gemini South F2 spectra, we observed with the JH grism and 0.72\arcsec~slit in place, yielding a wavelength range of $1.0-1.8\,\rm{\mu m}$. 
All observations were taken with a standard ABBA pattern for sky subtraction, and an A0V star was observed adjacent to the science exposures for both telluric corrections and flux calibration.  Data for both instruments was reduced in a standard way as described in \citet{Hsiao19}, and we refer the reader there for the details.

\newpage
\begin{center}
\startlongtable
\begin{deluxetable*}{cccc}
\tablecaption{DLT40 unfiltered light curves of SN~2017ahn. Magnitudes are calibrated to the $r$-band. \label{table:prompt}}
\tablehead{\colhead{Date} & \colhead{JD} & \colhead{phase} & \colhead{$Open$(err)} \\ 
\colhead{} & \colhead{} & \colhead{d} & \colhead{(mag)} } 
\startdata
20170205 & 2457789.74 & -2.52 & $>21.1$     \\
20170206 & 2457790.74 & -1.52 & $>21.0$     \\
20170207 & 2457791.73 & -0.53 & $>20.8$     \\
20170208 & 2457792.79 &  0.53 & 18.09(0.08) \\
20170208 & 2457792.84 &  0.58 & 18.07(0.08) \\
20170208 & 2457792.84 &  0.58 & 17.89(0.08) \\
20170209 & 2457793.84 &  1.58 & 16.82(0.06) \\
20170210 & 2457794.57 &  2.31 & 16.56(0.07) \\
20170211 & 2457795.56 &  3.30 & 15.90(0.05) \\
20170212 & 2457796.56 &  4.30 & 15.66(0.05) \\
\enddata
\tablecomments{
Data were obtained using the PROMPT5 $0.41\rm{m}$ telescope at the Cerro Tololo Inter-American Observatory, Cerro Pach\'on, Chile;
Phases refer to the epoch of the explosion. Table 1 is published in its entirety in the machine-readable format. A portion is shown here for guidance regarding its form and content.}
\end{deluxetable*}
\end{center}

\begin{center}
\startlongtable
\begin{deluxetable*}{ccccccc}
\tablecaption{$UBV$ light curves of SN~2017ahn \label{table:UBVphot}}
\tablehead{\colhead{Date} & \colhead{JD} & \colhead{phase} & \colhead{$U$(err)} & \colhead{$B$(err)} & \colhead{$V$(err)} & \colhead{Instrument} \\ 
\colhead{} & \colhead{} & \colhead{d} & \colhead{(mag)} & \colhead{(mag)} & \colhead{(mag)} & \colhead{}} 
\startdata
20170208 & 2457792.87 &   0.61 & \nodata     & 17.89(0.04) & 17.70(0.05) & 1m0-09  \\	
20170208 & 2457793.32 &   1.06 & 16.83(0.06) & 17.16(0.09) & 17.04(0.08) & 1m0-12  \\	
20170209 & 2457793.57 &   1.31 & 16.38(0.05) & 17.03(0.05) & 16.88(0.05) & 1m0-09  \\	
20170209 & 2457794.46 &   2.20 & 15.79(0.05) & 16.24(0.07) & 16.11(0.09) & 1m0-13  \\	
20170210 & 2457794.78 &   2.52 & 15.69(0.04) & 16.16(0.05) & 16.16(0.05) & 1m0-09  \\	
20170210 & 2457795.24 &   2.98 & 15.40(0.08) & 15.92(0.09) & 15.77(0.10) & 1m0-11  \\	
20170211 & 2457795.79 &   3.53 & 15.30(0.07) & 15.72(0.22) & 15.46(0.39) & 1m0-05  \\	
20170211 & 2457795.98 &   3.72 & 15.30(0.08) & 15.84(0.08) & 15.64(0.09) & 1m0-03  \\	
20170212 & 2457796.66 &   4.40 & 15.25(0.03) & 15.67(0.04) & 15.59(0.04) & 1m0-09  \\	
20170212 & 2457797.16 &   4.90 & 15.10(0.20) & \nodata	   & \nodata	 & 1m0-11  \\
\enddata
\tablecomments{
Las Cumbres Observatory 1m0-03, 1m0-11: node at Siding Spring, Australia; 1m0-04, 0m4-05, 1m0-05, 1m0-09: node at Cerro Tololo Inter-American Observatory, Chile; 1m0-10, 1m0-12, 1m0-13: node at South African Astronomical Observatory, South Africa. Phases refer to the epoch of the explosion. Table 2 is published in its entirety in the machine-readable format. A portion is shown here for guidance regarding its form and content.
}
\end{deluxetable*}
\end{center}

\begin{center}
\startlongtable
\begin{deluxetable*}{cccccccc}
\tablecaption{$griz$ light curves of SN~2017ahn \label{table:grizphot}}
\tablehead{\colhead{Date} & \colhead{JD} & \colhead{phase} & \colhead{$g$(err)} & \colhead{$r$(err)} & \colhead{$i$(err)} & \colhead{$z$(err)} & \colhead{Instrument} \\ 
\colhead{} & \colhead{} & \colhead{d} & \colhead{(mag)} & \colhead{(mag)} & \colhead{(mag)} & \colhead{(mag)} & \colhead{}} 
\startdata
20170208 & 2457792.880 &   0.62 & 17.81(0.02) & 17.88(0.04) & 18.03(0.06) &  \nodata    & 1m0-09 \\
20170208 & 2457793.220 &   0.96 & 17.42(0.14) &  \nodata    &  \nodata	  &  \nodata    & 1m0-11 \\
20170208 & 2457793.260 &   1.00 &  \nodata    &  \nodata    & 17.27(0.33) &  \nodata    & 0m4-03 \\
20170209 & 2457793.640 &   1.38 & 16.87(0.02) & 17.14(0.02) & 17.00(0.02) & 17.11(0.02) & GROND  \\
20170209 & 2457793.855 &   1.59 & 16.71(0.04) & 16.85(0.03) & 16.85(0.04) &  \nodata    & 1m0-05 \\
20170209 & 2457794.365 &   2.10 & 16.37(0.02) & 16.43(0.02) & 16.51(0.02) &  \nodata    & 1m0-13 \\
20170210 & 2457794.620 &   2.36 & 16.15(0.06) & 16.38(0.02) & 16.34(0.02) & 16.48(0.02) & GROND  \\
20170210 & 2457794.657 &   2.40 & 16.21(0.04) & 16.31(0.04) & 16.41(0.04) &  \nodata    & 1m0-05 \\
20170210 & 2457794.955 &   2.69 & 16.17(0.05) & 16.19(0.05) & 16.31(0.05) &  \nodata    & 1m0-11 \\
20170211 & 2457795.770 &   3.51 & 15.82(0.02) & 15.96(0.03) & 16.11(0.02) & 16.23(0.03) & GROND  \\
\enddata
\tablecomments{
GROND: MPG $2.2\,\rm{m}$ telescope with GROND at the ESO La Silla Observatory, Chile; Las Cumbres Observatory 1m0-03, 1m0-11: node at Siding Spring, Australia; 1m0-04, 0m4-05, 1m0-05, 1m0-09: node at Cerro Tololo Inter-American Observatory, Chile; 1m0-10, 1m0-12, 1m0-13: node at South African Astronomical Observatory, South Africa. Phases refer to the epoch of the explosion. Table 3 is published in its entirety in the machine-readable format. A portion is shown here for guidance regarding its form and content.}
\end{deluxetable*}
\end{center}

\begin{center}
\startlongtable
\begin{deluxetable*}{ccccccccc}
\tablecaption{UVOT light curves of SN~2017ahn \label{table:UVOTphot}}
\tablehead{\colhead{Date} & \colhead{JD} & \colhead{phase} & \colhead{$uvw2$(err)} & \colhead{$uvm2$(err)} & \colhead{$uvw12$(err)} & \colhead{$U$(err)} & \colhead{$B$(err)} & \colhead{$V$(err)} \\ 
           \colhead{}     & \colhead{}   & \colhead{(d)}   & \colhead{(mag)}       & \colhead{(mag)}       & \colhead{(mag)}        & \colhead{(mag)}    & \colhead{(mag)}    & \colhead{(mag)}  } 
\startdata
20170208 & 2457793.12 & 0.33  & \nodata     & \nodata     & \nodata     & \nodata     & \nodata     & 16.99(0.10) \\
20170208 & 2457793.13 & 0.34  & \nodata	    & 16.65(0.06) & \nodata     & \nodata     & \nodata	    & \nodata	  \\
20170208 & 2457793.15 & 0.35  & \nodata	    & \nodata	  & 16.49(0.07) & \nodata     & \nodata	    & \nodata	  \\
20170208 & 2457793.15 & 0.35  & \nodata	    & \nodata	  & \nodata     & 16.43(0.06) & \nodata	    & \nodata	  \\
20170208 & 2457793.15 & 0.35  & \nodata	    & \nodata	  & \nodata     & \nodata     & 17.42(0.06) & \nodata	  \\
20170208 & 2457793.15 & 0.36  & 16.62(0.08) & \nodata	  & \nodata     & \nodata     & \nodata	    & \nodata	  \\
20170209 & 2457793.75 & 0.96  & \nodata	    & \nodata	  & 15.64(0.06) & \nodata     & \nodata	    & \nodata	  \\
20170209 & 2457793.75 & 0.96  & \nodata	    & \nodata	  & \nodata     & \nodata     & 16.76(0.05) & \nodata	  \\
20170209 & 2457793.72 & 0.93  & \nodata	    & 15.84(0.06) & \nodata     & \nodata     & \nodata	    & \nodata	  \\
20170209 & 2457793.75 & 0.96  & \nodata	    & \nodata	  & \nodata     & \nodata     & \nodata	    & 16.58(0.07) \\
\enddata
\tablecomments{Data were obtained using the $0.3\,\rm{m}$ UV/optical Telescope (UVOT) on board of the {\it Swift} Gamma-Ray Burst Mission. Phases refer to the epoch of the explosion. Table 4 is published in its entirety in the machine-readable format. A portion is shown here for guidance regarding its form and content.}
\end{deluxetable*}
\end{center}

\begin{center}
\startlongtable
\begin{deluxetable*}{ccccccc}
\tablecaption{$JHK$ light curves of SN~2017ahn \label{table:JHKphot}}
\tablehead{\colhead{Date} & \colhead{JD} & \colhead{phase} & \colhead{$J$(err)} & \colhead{$H$(err)} & \colhead{$K$(err)}  \\ 
           \colhead{} & \colhead{}   & \colhead{(d)}      & \colhead{(mag)}    & \colhead{(mag)}    & \colhead{(mag)} } 
\startdata
20170209 & 2457793.640 &  1.38 & 16.30(0.04) & 16.16(0.04) & 16.11(0.05) \\
20170210 & 2457794.620 &  2.36 & 15.93(0.05) & 15.71(0.04) & 15.61(0.05) \\
20170211 & 2457795.770 &  3.51 & 15.53(0.03) & 15.39(0.06) & 15.25(0.04) \\
20170212 & 2457796.760 &  4.50 & 15.33(0.05) & 15.21(0.05) & 15.11(0.05) \\
20170213 & 2457797.680 &  5.42 & 15.11(0.08) & 15.10(0.04) & 15.05(0.05) \\
20170214 & 2457798.580 &  6.32 & 15.13(0.04) & 15.12(0.04) & 14.99(0.04) \\
20170215 & 2457799.610 &  7.35 & 15.11(0.04) & 14.95(0.05) & 14.83(0.04) \\
20170218 & 2457802.660 & 10.40 & 14.74(0.04) & 14.64(0.06) & 14.72(0.04) \\
20170222 & 2457806.730 & 14.47 & 14.72(0.04) & 14.51(0.06) &  \nodata    \\
\enddata
\tablecomments{Data were obtained using the MPG $2.2\,\rm{m}$ telescope with GROND, at the ESO La Silla Observatory, Chile. Phases refer to the epoch of the explosion.}
\end{deluxetable*}
\end{center}

\begin{center}
\begin{deluxetable*}{cccccccc}
\tablecaption{Log of the spectroscopic observations of SN~2017ahn \label{table:speclog}}
\tablehead{\colhead{Date} & \colhead{JD} & \colhead{Phase} & \colhead{Instrumental setup} & \colhead{Grism/Grating} & \colhead{Spectral range} & \colhead{Exposure time} & \colhead{Resolution} \\ 
\colhead{} & \colhead{} & \colhead{(d)} & \colhead{} & \colhead{} & \colhead{(\AA)} & \colhead{(s)} & \colhead{(\AA)} } 
\startdata
20170209 & 2457793.70 & $+1.4$  & GeminiS+FLAMINGOS2 & JH              & $10000-18000$ & $2400$ & 14.0 \\
20170209 & 2457793.97 & $+1.7$  & FTN+FLOYDS    & $235\,\rm{l/mm}$    & $3500-9000$  & $3600$  & 15.0 \\
20170209 & 2457794.37 & $+2.1$  & SALT+RSS      & PG0900              & $3500-9000$  & $2500$  & 5.6 \\
20170210 & 2457794.96 & $+2.7$  & FTS+FLOYDS    & $235\,\rm{l/mm}$    & $5000-10000$ & $3600$ & 14.3  \\
20170212 & 2457796.82 & $+4.6$  & SOAR+GOODMAN  & SYZY\_400           & $3500-8000$  & $900$  &  6.5 \\
20170213 & 2457798.35 & $+6.1$  & SALT+RSS      & PG0900              & $3500-9000$  & $2600$  & 5.6 \\
20170214 & 2457798.97 & $+6.7$  & FTS+FLOYDS    & $235\,\rm{l/mm}$    & $5000-10000$ & $3600$ & 16.9  \\
20170215 & 2457799.92 & $+7.7$  & FTS+FLOYDS    & $235\,\rm{l/mm}$    & $5000-10000$ & $3600$  & 17.0 \\
20170216 & 2457801.10 & $+8.8$  & FTS+FLOYDS    & $235\,\rm{l/mm}$    & $5000-10000$ & $3600$  & 17.2 \\
20170217 & 2457801.91 & $+9.6$  & IRTF+Spex     & ShortXD             & $8000-24000$ & $2400$  & 11.7 \\
20170219 & 2457804.19 & $+11.9$ & FTS+FLOYDS    & $235\,\rm{l/mm}$    & $5000-10000$ & $3600$ & 19.0  \\
20170222 & 2457807.24 & $+15$   & FTS+FLOYDS    & $235\,\rm{l/mm}$    & $5000-10000$ & $3600$  & 17.7 \\
20170305 & 2457817.91 & $+26$   & FTN+FLOYDS    & $235\,\rm{l/mm}$    & $5000-10000$ & $3600$  & 14.8 \\
20170311 & 2457823.89 & $+32$   & IRTF+Spex     & ShortXD             & $8000-24000$ & $3000$  & 11.7 \\
20170315 & 2457828.53 & $+36$   & SALT+RSS      & PG0900              & $3500-9000$  & $2230$ & 5.6  \\
20170316 & 2457828.90 & $+37$   & FTN+FLOYDS    & $235\,\rm{l/mm}$    & $5000-9000$  & $3600$  & 16.5 \\
20170324 & 2457836.83 & $+45$   & FTN+FLOYDS    & $235\,\rm{l/mm}$    & $5000-9000$  & $3600$  & 15.8 \\
20170326 & 2457838.69 & $+46$   & Baade+FIRE       & LDPrism             & $8500-24000$ & $2282$ &  24.0\\
20170327 & 2457840.48 & $+48$   & SOAR+GOODMAN  & SYZY\_400           & $3500-8000$  & $ 900$  & 5.7 \\
20170330 & 2457842.92 & $+51$   & FTS+FLOYDS    & $235\,\rm{l/mm}$    & $5000-9000$  & $3600$  & 19.0 \\
20170405 & 2457848.80 & $+57$   & FTN+FLOYDS    & $235\,\rm{l/mm}$    & $5000-9000$  & $3600$  & 14.9 \\
20170414 & 2457857.69 & $+65$   & Baade+FIRE     & LDPrism             & $8500-24000$ & $2028$  &  24.0\\
20170420 & 2457864.00 & $+72$   & FTS+FLOYDS    & $235\,\rm{l/mm}$    & $5000-9000$  & $3600$  & 19.3 \\
20170505 & 2457879.01 & $+87$   & FTS+FLOYDS    & $235\,\rm{l/mm}$    & $5000-9000$  & $3600$  & 18.7 \\
20170516 & 2457889.94 & $+98$   & FTS+FLOYDS    & $235\,\rm{l/mm}$    & $5000-9000$  & $3600$  & 19.8 \\
20170527 & 2457900.86 & $+109$  & FTS+FLOYDS    & $235\,\rm{l/mm}$    & $5000-9000$  & $3600$ & 19.8 \\
20170530 & 2457903.77 & $+112$  & KECK+LRIS     & $600/4000+400/8500$ & $3500-10000$ & $180+180$ & 6.5 \\
20180526 & 2458264.63 & $+472$ & VLT+MUSE        & VPH                                & $4600-9200$ & $4\times600$ & 3.0 \\
\enddata
\tablecomments{FTN: $2\,\rm{m}$ Faulkes Telescope North, Las Cumbres Observatory node at the Haleakala Observatory, Hawaii; FTS: $2\,\rm{m}$ Faulkes Telescope South, Las Cumbres Observatory node at the Siding Spring Observatory, Australia; SALT: $10\,\rm{m}$ Southern Africa Large Telescope at the South African Astronomical Observatory (SAAO), Sutherland, South Africa; SOAR: SOuthern Astrophysical Research telescope and GeminiS: $8\,\rm{m}$ Gemini South telescope, both at the Cerro Tololo Inter-American Observatory, Cerro Pach\'on, Chile; BAADE: $\,\rm{m}$ Magellan 1 - Baade Telescope located at the Las Campanas Observatory of the Carnegie Institution for Science, Cerro Las Campanas, Chile; KECK: $10\,\rm{m}$ Keck I telescope, at the Mauna Kea Observatory, Hawaii -- U.S. MUSE: $8.2\,\rm{m}$ Very Large Telescope, European Southern Observatory, Cerro Paranal, Chile.}
\end{deluxetable*}
\end{center}

\end{document}